\documentclass[%
 reprint,
 amsmath,amssymb,
 aps, 
]{revtex4-2}

\usepackage[dvipdfmx]{graphicx} % Include figure files
\usepackage{dcolumn}% Align table columns on decimal point
\usepackage{bm}% bold math
\usepackage{color} %change font color
\usepackage{braket} %braket
\usepackage{ulem}
\usepackage{multirow} %merge a row of a Table

\usepackage{mathrsfs}                    %capital letters
\usepackage{tensor}                      %tensor
\usepackage{cancel}                      %Feynman slash

\newcommand{\dd}{\mathrm{d}} %roman d
\newcommand{\tr}{\mathrm{tr}} %roman trace
\newcommand{\Det}{\mathrm{Det}} %functional determinant

\begin{document}

\preprint{}

\title{
Investigations of symmetry breakings by multiple condensates \\ on exotic chiral condensed phases
}

\author{Joshua Murakami}
\affiliation{Graduate School of Integrated Arts and Sciences, Kochi University, Kochi 780-8520, Japan}
\author{Kentaro Hayashi}
\affiliation{Graduate School of Integrated Arts and Sciences, Kochi University, Kochi 780-8520, Japan}
\author{Yasuhiko Tsue}
\affiliation{Department of Mathematics and Physics, Faculty of Science and Technology, Kochi University, Kochi 780-8520, Japan}

\date{\today}

\begin{abstract}
The Nambu-Goldstone modes on the exotic chiral condensed phase with chiral and tensor-type quark-antiquark condensates are investigated by using the two-point vertex functions.
It is shown that one of the Nambu-Goldstone modes appears as a result of meson mixing.
As is well known, another method to find the Nambu-Goldstone modes is given by the use of the algebraic commutation relations between broken generators and massless modes obtained through the spontaneous symmetry breaking.
This method is adopted to the cases of the chiral symmetry breakings due to the tensor-type condensate and the inhomogeneous chiral condensate.
The result obtained by the use of the meson two-point vertex functions is obviously reproduced in the case of the tensor-type condensate. Furthermore, we investigate the general rules for determining the broken symmetries and the Nambu-Goldstone modes algebraically.
As examples, the symmetry breaking pattern and the Nambu-Goldstone modes due to the tensor-type condensate or the inhomogeneous chiral condensate are shown by adopting the general rules developed in this paper in the algebraic method.
\end{abstract}

\pacs{Valid PACS appear here}% PACS, the Physics and Astronomy
\maketitle
%%%%%%%%%%%%%%%%%%%%%%%%%%%%%%%%%%%%%
%%%%%%%%%%%%%%%%%%%%%%%%%%%%%%%%%%%%%
\section{\label{sec:level1}Introduction}

One of the recent interests of physics in the world governed by the quantum chromodynamics (QCD) may be to clarify the phase diagram of the quark-gluon and/or hadron world \cite{FH}. 
Of course, in the region with zero temperature and small baryon chemical potential, the chiral symmetry broken phase is realized, in which the chiral symmetry is dynamically broken due to the quark-antiquark condensate \cite{NJL,MN,Higashijima,Kugo,RW}.
On the other hand, at low temperature and finite density, it has widely been pointed out that there exists the possibility of various phases, such as the color superconducting phase \cite{CS,IB,ASRS}, the quarkyonic phase \cite{MP}, and the exotic chiral broken phases.
Especially, as for the exotic chiral broken phases, a quark spin polarized phase due to a pseudovector-type quark-antiquark condensate \cite{Maedan,Morimoto}, the chiral symmetry broken phase with a tensor-type quark-antiquark condensate \cite{YT2012}, or the inhomogeneous chiral condensate \cite{NT}have been investigated by using the Nambu–Jona-Lasinio (NJL) model \cite{NJL} as an effective model of QCD.

It is well known that the massless Nambu-Goldstone (NG) mode has to appear when the continuous global symmetry is broken \cite{G,GSW}. 
In the chiral symmetry broken phase owing to the scalar-type quark-antiquark chiral condensate, namely so-called chiral condensate, the NG modes are pions that take responsibility for the low energy hadron dynamics.
Thus, it may be important to find the NG modes on the various symmetry broken phases.
In previous papers, a possibility of the existence of the pseudovector-type quark-antiquark condensed phase in the quark matter is investigated that leads to the chiral symmetry breaking\cite{Maedan,Morimoto}.  
In Ref.\ \cite{Hayashi:2023apo}, the massless NG modes on the pseudovector condensate due to the pseudovector interaction in the NJL model are investigated directly, and it has been shown that two NG modes appear through the mesons mode mixing if both the chiral condensate and the pseudovector-type condensate coexist.

In this paper, first, it is shown that the NG mode also appears with the meson mixing by means of the two-point vertex function developed in the previous paper \cite{Hayashi:2023apo} and/or the algebraic commutation relations between broken generators and massless modes when both the chiral condensate and the tensor-type condensate coexist.
Second, by using the algebraic method, the NG mode is investigated in the case of the chiral symmetry breaking due to the inhomogeneous chiral condensate.
In this case, the translational and rotational symmetries are also broken.
It is found that the result obtained by Low-Manohar is reproduced \cite{Low:2002sbs}. 
In addition, it also shows the coupling of internal and translational symmetries as mentioned by Kobayashi and Nitta \cite{Kobayashi:2014nrn}.
Third, the general discussion is given in the case of the symmetry breaking due to the multiple condensates. 
Also, it is found that the mode mixing and/or symmetry generator mixing are seen in the general framework. 
Finally, the results obtained in the cases that both the chiral and the tensor condensates coexist or the inhomogeneous chiral condensate appears are reproduced with the help of the general framework.

In general, the dispersion law for small momentum has been indicated clearly by Nambu \cite{Nambu}. 
Further, the counting rule for NG modes has been revealed in \cite{Hidaka}, and a simple criterion to determine what redundancies exist among NG modes in a given system has been presented in \cite{WM}. 
In our treatment, it is shown that the reduction of symmetry generators by a linear combination leads to the reduction of NG modes as was discussed in \cite{WM}. 
Further, it is shown that there is a case in which the NG modes appear through the linear combinations of elementary fields included originally in a model.

This paper is organized as follows:
In Sec. \ref{sec:level2}, the massless modes are investigated when the chiral condensate and the tensor-type quark-antiquark condensate (tensor condensate) coexist, following the previous paper \cite{Hayashi:2023apo} in the NJL model with the tensor interaction between quarks. 
The detailed calculations to the gap equations and the two-point vertex functions are shown in Appendixes \ref{sec:appendixA} and \ref{sec:appendixB}. 
In Sec. \ref{sec:level3}, the spontaneous symmetry breaking is considered in terms of commutation relations following the previous paper \cite{Hayashi:2023apo}. 
In Sec. \ref{sec:level3-1}, we revisit the massless modes discussed in Sec. \ref{sec:level2}, demonstrating that the same results are obtained using these commutation relations. Moreover, we also discuss spatial symmetry and conclude that NG modes caused by chiral symmetry breaking do not mix with NG modes caused by spatial rotation symmetry breaking. In Sec. \ref{sec:level3-2}, we assume the dual chiral density wave as an inhomegeneous chiral condensate and investigate the massless modes. In that case, it is shown that the spatial translation symmetry, the spatial rotation symmetry and the internal symmetry are not independent as was mentioned in Refs. \cite{Low:2002sbs, Kobayashi:2014nrn}.
In Sec. \ref{sec:level4}, we present the general theory of one symmetry broken by two or more condensates and the general theory of two symmetries broken by two condensates in a similar manner to the discussion in Sec. \ref{sec:level3}. 
We show that the results of the massless modes in the case of one symmetry broken by two condensates developed in Secs. \ref{sec:level2} and \ref{sec:level3-2} can be correctly obtained within the framework of the general theory in Sec. \ref{sec:level4-1}.
Also, it is shown that the massless modes in the case of two symmetries broken by two condensates developed in Sec. \ref{sec:level3-2} are given in the general theory in Sec. \ref{sec:level4-3}. 
Section \ref{sec:level5} is devoted to a summary.

%%%%%%%%%%%%%%%%%%%%%%%%%%%%%%%%%%%%%
%%%%%%%%%%%%%%%%%%%%%%%%%%%%%%%%%%%%%

\section{\label{sec:level2}The massless modes on the chiral and tensor condensates}

In this section, the massless modes due to the chiral symmetry breaking caused by both the chiral condensate and the tensor-type quark-antiquark condensate (tensor condensate) are investigated in the NJL model by searching for the zero point in the meson propagators. In this section, we search the massless modes by means of the two-point vertex functions.

We introduce the NJL model with tensor-type interaction, which retains the chiral symmetry:
\begin{align}
\mathscr{L}_{ \mathrm{NJL, T} } &= \mathscr{L}_{\mathrm{NJL}} + \mathscr{L}_{\mathrm{T}}, 
\label{eq:L_NJL_T} \\ 
\notag \\ 
\mathscr{L}_{\mathrm{NJL}} &= \bar{\psi} i \gamma^{\mu} \partial_{\mu} \psi 
 + G_{S} \left[ (\bar{\psi} \psi)^{2} + (\bar{\psi} i \gamma_{5} \bm{\tau} \psi )^{2} \right] , 
\label{eq:L_NJL} \\ 
\notag \\ 
\mathscr{L}_{\mathrm{T}} &= G_{T} \left[ (\bar{\psi} \gamma^{\mu} \gamma^{\nu} \bm{\tau} \psi)^{2} + (\bar{\psi} i \gamma_{5} \gamma^{\mu} \gamma^{\nu} \psi )^{2} \right]
\label{eq:L_T}
\end{align}
with $\mu \neq \nu$ in \eqref{eq:L_T}, where $\bm{\tau}$ represents the isospin matrices of flavor $su(2)-$symmetry. 
Additionally, regarding the tensor interaction terms $\mathscr{L}_T$,  while it is concise to use the Hermitian antisymmetric tensor $\sigma_{\mu\nu}$ as in the paper \cite{JaminonArriola}, we use the expression in \eqref{eq:L_T}, to clearly demonstrate the preservation of chiral symmetry with $\gamma_{5}$. 
The generating functional $Z$ is written as
\begin{equation}
Z = \int \mathscr{D} \psi \mathscr{D} \bar{\psi} \, \exp \left(i \int \dd^{4} x \, \mathscr{L}_{ \mathrm{NJL, T} } \right). 
\label{eq:Z_1}
\end{equation}
Here, using the auxiliary field method, we introduce the functional Gaussian integrals as follows:
\begin{align}
   \begin{split}
1 &= \int \mathscr{D} \sigma' \mathscr{D} \bm{\pi}' \, \exp \left( i \int \dd^{4}x \left[- \frac{1}{G_{S}} \left( \sigma'^{2} + \bm{\pi}'^{2} \right) \right]  \right), 
\\ 
1 &= \int \mathscr{D} \tensor{{\bm{t}'}}{^\mu^\nu} \mathscr{D} \tensor{{\varpi'}}{^\mu^\nu}
\\ 
& \hspace{10pt} \times \exp \left( i \int \dd^{4}x \left[- \frac{1}{G_{T}} \left( \tensor{{\bm{t}'}}{^\mu^\nu} \tensor{{\bm{t}'}}{_\mu_\nu}  + \tensor{{\varpi'}}{^\mu^\nu}  \tensor{{\varpi'}}{_\mu_\nu} \right)  \right]  \right).
   \end{split}
\label{eq:FGI}
\end{align}
Next, we insert the above units into the generating functional $Z$ in \eqref{eq:Z_1} and replace $\sigma', \bm{\pi}', \tensor{{\bm{t}'}}{^\mu^\nu}$ and $\tensor{{\varpi'}}{^\mu^\nu}$ into $\sigma, \bm{\pi}, \tensor{ \bm{t} }{^\mu^\nu}$ and $\tensor{ \varpi }{^\mu^\nu}$ as
\begin{align}
   \begin{split}
\sigma' &= \sigma + G_{S} (\bar{\psi} \psi), 
\\ 
\bm{\pi}' &= \bm{\pi} + G_{S} (\bar{\psi} i \gamma_{5} \bm{\tau} \psi), 
\\ 
\tensor{{ \bm{t}'} }{^\mu^\nu} &= \tensor{ \bm{t} }{^\mu^\nu} + G_{T} \left( \bar{\psi} \gamma^{\mu} \gamma^{\nu} \bm{\tau} \psi \right), 
\\ 
\tensor{{\varpi'}}{^\mu^\nu} &= \tensor{ \varpi }{^\mu^\nu} + G_{T} \left(\bar{\psi} i \gamma_{5} \gamma^{\mu} \gamma^{\nu} \psi \right).
   \end{split}
\label{eq:replacing}
\end{align}
The generating functional $Z$, then, becomes
\begin{equation}
Z = \int \mathscr{D} \psi \mathscr{D} \bar{\psi} \int \mathscr{D} \sigma  \mathscr{D} \bm{\pi} \mathscr{D} \tensor{ \bm{t} }{^\mu ^\nu} \mathscr{D} \tensor{ \varpi }{^\mu^\nu} \, \exp \left(i \int \dd^{4} x \, \tilde{ \mathscr{L} } \right), 
\label{eq:Z_2}
\end{equation}
where
\begin{align}
\tilde{ \mathscr{L} } 
&= 
\bar{\psi} \left( 
i \gamma^{\mu} \partial_{\mu} - 2 \sigma - 2 i \gamma_{5} \bm{\tau} \cdot \bm{\pi} \right.
\nonumber \\ 
& \hspace{10pt} 
\left. - 2 \gamma^{\mu} \gamma^{\nu} \bm{\tau} \cdot \tensor{ \bm{t} }{_\mu_\nu}  - 2 i \gamma_{5} \gamma^{\mu} \gamma^{\nu} \tensor{ \varpi }{_\mu_\nu} \right) \psi 
\nonumber \\ 
& \hspace{20pt} 
- \frac{1}{G_{S}} \left( \sigma^{2} + \bm{\pi}^{2} \right) 
- \frac{1}{G_{T}} \left( \tensor{ \bm{t} }{^\mu^\nu} \cdot \tensor{ \bm{t} }{_\mu_\nu} + \tensor{ \varpi }{^\mu^\nu} \tensor{ \varpi }{_\mu_\nu} \right).
\label{eq:L_tilde}
\end{align}
Here, from $\delta \tilde{ \mathscr{L} }/ \delta \alpha = 0 \hspace{4pt} (\alpha = \sigma, \bm{\pi}, \tensor{ \bm{t} }{^\mu^\nu}, \tensor{ \varpi }{^\mu^\nu})$, we obtain
\begin{align}
   \begin{split}
\sigma &= - G_{S} \left(\bar{\psi} \psi \right) , 
\\ 
\bm{\pi} &= -G_{S} \left(\bar{\psi} i \gamma_{5} \bm{\tau} \psi \right) , 
\\ 
\tensor{\bm{t}}{_\mu_\nu} &= - G_{T} \left(\bar{\psi} \gamma_{\mu} \gamma_{\nu} \bm{\tau} \psi \right),  
\\ 
\tensor{ \varpi }{_\mu_\nu} &= - G_{T} \left( \bar{\psi} i \gamma_{5} \gamma_{\mu} \gamma_{\nu} \psi \right). 
   \end{split}
\label{eq:kinetic_eq}
\end{align} 
By integrating over the $\psi$ and $\bar{\psi}$ in \eqref{eq:Z_2}, the generating functional $Z$ and the effective action $\Gamma$ are derived as
\begin{align}
Z 
&= \int \mathscr{D} \sigma \mathscr{D} \bm{\pi} \mathscr{D} \tensor{\bm{t}}{^\mu^\nu} \mathscr{D} \tensor{ \varpi }{^\mu^\nu} \exp (i \Gamma) , 
\nonumber \\ 
     \begin{split}
\Gamma 
&= \int \dd^{4} x \left[ - \frac{1}{G_{S}} (\sigma^{2} + \bm{\pi}^{2}) - \frac{1}{G_{T}} (\tensor{\bm{t}}{^\mu^\nu} \cdot \tensor{\bm{t}}{_\mu_\nu} + \tensor{ \varpi }{^\mu^\nu} \tensor{ \varpi }{_\mu_\nu}) \right] 
\\ 
& \hspace{45pt} - i \ln \Det \left( i \gamma^{\mu} \partial_{\mu} - 2 \sigma - 2 i \gamma_{5} \bm{\tau} \cdot \bm{\pi} \right.
\\ 
& \hspace{90pt} \left. - 2 \gamma^{\mu} \gamma^{\nu} \bm{\tau} \cdot \tensor{ \bm{t} }{_\mu_\nu}  - 2 i \gamma_{5} \gamma^{\mu} \gamma^{\nu}  \tensor{ \varpi }{_\mu_\nu}  \right). 
     \end{split}
\label{eq:EA}
\end{align}
Here, Det means a functional determinant, which is a determinant with respect to space-time coordinates and all other field indices.
We should note that the tensor mode $\tensor{\bm{t}}{^1^2}$ is written as
\begin{equation}
\tensor{\bm{t}}{^1^2} 
= - G_{T} \bar{\psi} \gamma^{1} \gamma^{2} \bm{\tau} \psi 
= i G_{T} \bar{\psi} \Sigma_{3} \bm{\tau} \psi. 
\label{eq:tensor_mode}
\end{equation}
Here, we adopt the following Dirac representation for the gamma matrices:
\begin{gather}
   \begin{split}
   \gamma^{\mu} = (\gamma^{0}, \bm{\gamma})
   , \hspace{5pt}
   \gamma^{0} = 
   \begin{pmatrix}
   1 & 0 \\ 
   0 & -1
   \end{pmatrix}
   , \hspace{5pt}
   \bm{\gamma} = 
   \begin{pmatrix}
   0 & \bm{\sigma} \\ 
   -\bm{\sigma} & 0
   \end{pmatrix}, 
\\ 
   \gamma^{5} 
   = 
   \begin{pmatrix}
   0 & \bm{1} \\ 
   \bm{1} & 0
   \end{pmatrix}
   , \hspace{5pt}
   \gamma^{1} \gamma^{2} 
   = -i \Sigma_{3} 
   = -i 
   \begin{pmatrix}
   \sigma_3 & 0 \\ 
   0 & \sigma_3
   \end{pmatrix},
\end{split} 
\end{gather} 
where $\bm{\sigma}$ and $\sigma_{3}$ represent the Pauli spin matrices and third component. 
Thus, $\Sigma_{3}$ is the third component of the spin matrices.

In the following, we assume the existence of the chiral condensate $\sigma_{0} = \langle \sigma \rangle$ and the tensor condensate $\tilde{t}_{0} = \langle t_{i = 3}^{\mu = 1, \nu = 2} \rangle$ with $\mu = 1, \nu = 2$, and the third component of the isospin:
\begin{align}
\sigma_{0} 
&= \langle \sigma \rangle
= - G_{S} \langle \bar{\psi} \psi \rangle, 
\\ 
\tilde{t}_{0} 
&= \langle t_{i = 3}^{\mu = 1, \nu = 2} \rangle
= - G_{T} \langle \bar{\psi} \gamma^{1} \gamma^{2} \tau_{3} \psi \rangle 
= i G_{T} \langle \bar{\psi} \Sigma_{3} \tau_{3} \psi \rangle .
\end{align} 
Here, we conventionally chose the third isospin component in the condensate. 
Since $\tilde{t}_{0} $ is purely imaginary, we consider $t_{0}$ as follows:
\begin{equation}
t_{0} = -i \tilde{t}_{0} = G_{T} \langle \bar{\psi} \Sigma_{3} \tau_{3} \psi \rangle. 
\end{equation}
Thus, $t_0$  may be regarded as a spin polarized condensate. 
Under these quark-antiquark condensates, the effective potential $V(\sigma_{0}, t_{0})$ can be derived from the effective action
\begin{align}
\Gamma (\sigma_{0}, t_{0}) 
&=
- V (\sigma_{0}, t_{0}) \int \dd^{4} x , 
\label{eq:effective_action}
\\ 
V (\sigma_{0}, t_{0})
&= 
\frac{1}{G_{S}} \sigma_{0}^{2} - \frac{2}{G_{T}} t_{0}^{2} 
\nonumber \\ & \hspace{10pt}
- \int \frac{\dd^{4} k}{i (2 \pi)^{4}} \ln \det \left[ \cancel{k} - 2 \sigma_{0} - 4 i t_{0} \gamma^{1} \gamma^{2} \tau_{3} \right]
\nonumber \\ 
&= 
\frac{1}{G_{S}} \sigma_{0}^{2} - \frac{2}{G_{T}} t_{0}^{2} 
\nonumber \\ & \hspace{10pt}
- \int \frac{\dd^{4} k}{i (2 \pi)^{4}} \tr \ln \left[ \cancel{k} - 2 \sigma_{0} - 4 i t_{0} \gamma^{1} \gamma^{2} \tau_{3} \right], 
\label{eq:effective_potential}
\end{align}
where trace is taken by the Dirac gamma matrices, isospin space, and color space. 
Here, det means a determinant with respect to field indices other than spacetime coordinates.

From \eqref{eq:effective_potential}, the condensates $\sigma_{0}$ and $t_{0}$ are determined by the following gap equations:
\begin{align}
\frac{\partial V}{\partial \sigma_{0}} 
&= \frac{2}{G_{S}} \sigma_{0} - \int \frac{\dd^{4} k}{i (2\pi)^{4}} 
\notag \\ 
& \hspace{12pt} \times
\tr \left[ \frac{1}{ \cancel{k} - 2 \sigma_{0} - 4 i t_{0} \gamma^{1} \gamma^{2} \tau_{3} } \left( - 2 \right) \right] 
= 0, 
\label{eq:gap_chiral1}
\\ 
\frac{\partial V}{\partial t_{0}} 
&= - \frac{4}{G_{T}} t_{0}  - \int \frac{\dd^{4} k}{i (2\pi)^{4}}
\notag \\ 
& \hspace{12pt} \times
\tr \left[ \frac{1}{ \cancel{k} - 2 \sigma_{0} - 4 i t_{0} \gamma^{1} \gamma^{2} \tau_{3} } \left( - 4 i \gamma^{1} \gamma^{2} \tau_{3} \right) \right] 
= 0. 
\label{eq:gap_tensor1}
\end{align}
Through the calculations related to the gamma matrices provided in Appendix \ref{sec:appendixA},  we obtain the following expressions:
\begin{align}
&\sigma_{0} \left(  1 + 48 G_{S} \int \frac{\dd^{4} k}{i (2\pi)^{4}}  \right.
\notag \\ 
&\hspace{12pt} \times \left.
\frac{ k^{2} - 4 \sigma_{0}^{2} + 16 t_{0}^{2} }{ \displaystyle (k^{2} - 4 \sigma_{0}^{2} - 16 t_{0}^{2})^{2} - 64 t_{0}^{2} \left( k_{1}^{2} + k_{2}^{2} + 4 \sigma_{0}^{2} \right) }  \right)
= 0, 
\label{eq:gap_chiral2}
\\ 
&t_{0} \left( 1 - 96 G_{T} \int \frac{\dd^{4} k}{i (2\pi)^{4}} \right.
\notag \\ 
&\hspace{12pt} \times \left.
\frac{ ( k^{2} - 4 \sigma_{0}^{2} - 16 t_{0}^{2} ) + 2 ( k_{1}^{2} + k_{2}^{2} + 4 \sigma_{0}^{2} )}{ \displaystyle  (k^{2} - 4 \sigma_{0}^{2} - 16 t_{0}^{2})^{2} - 64 t_{0}^{2} \left( k_{1}^{2} + k_{2}^{2} + 4 \sigma_{0}^{2} \right) } \right) 
=0. 
\label{eq:gap_tensor2}
\end{align}
Here, $k^{2} = k_{0}^{2} - \bm{k}^{2}$, and $k_{1}$ and $k_{2}$ are the first and second components of the momentum vector $\bm{k}$.

To consider the coexistence of both the condensates $\sigma_{0}$ and $t_{0}$, let $\sigma_{0} \neq 0$ and $t_{0} \neq 0$. In this case,  Eqs. \eqref{eq:gap_chiral2} and \eqref{eq:gap_tensor2} respectively, become
\begin{align}
&1 + 48 G_{S} \int \frac{\dd^{4} k}{i (2\pi)^{4}} 
\notag \\ 
&\hspace{12pt} \times 
\frac{ k^{2} - 4 \sigma_{0}^{2} + 16 t_{0}^{2} }{ \displaystyle (k^{2} - 4 \sigma_{0}^{2} - 16 t_{0}^{2})^{2} - 64 t_{0}^{2} \left( k_{1}^{2} + k_{2}^{2} + 4 \sigma_{0}^{2} \right) } 
= 0, 
\label{eq:gap_chiral_coexist}
\\ 
&1 - 96 G_{T} \int \frac{\dd^{4} k}{i (2\pi)^{4}} 
\notag \\ 
&\hspace{12pt} \times 
\frac{ ( k^{2} - 4 \sigma_{0}^{2} - 16 t_{0}^{2} ) + 2 ( k_{1}^{2} + k_{2}^{2} + 4 \sigma_{0}^{2} )}{ \displaystyle  (k^{2} - 4 \sigma_{0}^{2} - 16 t_{0}^{2})^{2} - 64 t_{0}^{2} \left( k_{1}^{2} + k_{2}^{2} + 4 \sigma_{0}^{2} \right) }
=0. 
\label{eq:gap_tensor_coexist}
\end{align}
These are the gap equations for $\sigma_0\neq 0$ and $t_0\neq 0$. 

Next, we investigate the massless NG modes in the case where condensates $\sigma_{0}$ and $t_{0}$ coexist. The meson propagator $\Delta_{\alpha}^{-1} (x, y)$ can be derived from the effective action as
\begin{equation}
\Delta_{\alpha}^{-1} (x, y) = - \frac{\delta^{2} \Gamma}{\delta \alpha (x) \delta \alpha (y)}, 
\label{eq:meson_propagator}
\end{equation}
where $\alpha (x) = \sigma (x), \bm{\pi} (x),\bm{t}^{\mu\nu}(x) , \varpi^{\mu\nu}(x)$. In the momentum representation, the two-point vertex function $\Gamma_{\alpha\beta} (p)$ can be expressed as follows:
\begin{equation}
\Gamma_{\alpha\beta} (p) (2 \pi)^{4} \delta^{4} (p + q) 
= \left. 
\frac{\delta^{2} \Gamma}{\delta \alpha (p) \delta \beta (q)}
\right|_{\substack{\sigma = \sigma_{0}, t_{3}^{12} = i t_{0}, \\ \bm{\pi} = \varpi^{\mu\nu} = \bm{t}_{i \neq 3} = t_{i = 3}^{\mu \neq 1, \nu \neq 2} = 0}}. 
\label{eq:tpvf}
\end{equation} 
Here, $p, q$ represent the four-momentum, and the two-point vertex function $\Gamma_{\alpha\beta} (p)$ satisfies
\begin{equation}
\Gamma_{\alpha\alpha} (p) = \Gamma_{\alpha} (p) = - \Delta_{\alpha}^{-1} (p). 
\label{eq:tpvf_and_propagator}
\end{equation}
When $p^{2}$ is small, $\Gamma$ behaves according to the following expression:
\begin{equation}
\Gamma_{\alpha} (p) = - \Delta_{\alpha}^{-1} (p) \approx Z_{\alpha}^{-1} (m_{\alpha}^{2} - p^{2} ) + \cdots ,
\label{eq:tpvf_and_mass}
\end{equation}
where $m_{\alpha}$ is the meson masses. 
Although it is generally necessary to include the wave function renormalization constant $Z_\alpha$, it is not necessary to evaluate it explicitly because we focus on the massless modes only in this paper away from the values of masses of the Higgs modes.

We calculate the two-point vertex functions as follows:
\begin{align}
   \begin{split}
\Gamma_{\pi_{i}} &= \Gamma_{ \pi_{i} \pi_{i}}, \hspace{8pt} (i = 1, 2) 
\\ 
\Gamma_{\pi_{3}} &= \Gamma_{\pi_{3}\pi_{3}}, 
\\ 
\Gamma^{\mu\nu}_{\pi_{3} \varpi} &= \Gamma_{\pi_{3} \varpi^{\mu\nu}},
\\ 
\Gamma^{\mu\nu}_{\varpi \pi_{3}} &= \Gamma_{\varpi^{\mu\nu} \pi_{3}}, 
\\ 
\Gamma^{\mu\nu, \rho \sigma}_{\varpi} &= \Gamma_{ \varpi^{\mu\nu} \varpi^{\rho\sigma}}, 
\\ 
\Gamma^{\mu\nu, \rho \sigma}_{t_{i}} &= \Gamma_{ t_{i}^{\mu\nu} t_{i}^{\rho\sigma}}. \hspace{8pt} (i = 1, 2) 
   \end{split}
\label{eq:tpvf_mesons}
\end{align}
Here, since $\Gamma_{\pi_{i}}(p \to 0) = \Gamma^{12, 12}_{t_{i}}(p \rightarrow 0) = 0$ with $i = 1$ and $i = 2$,  $\pi_{1}, \pi_{2}, t_{1}^{12}$, and $t_{2}^{12}$ are massless modes. 
This can be confirmed as follows.
By expressing the effective action \eqref{eq:effective_action} in momentum space and substituting into \eqref{eq:tpvf}, we obtain the following expressions: 
\begin{align}
\Gamma_{\pi_{i = 1, 2}} (p) 
&= - \frac{2}{G_{S}} - \int \frac{\dd^{4} k}{i (2 \pi)^{4}} 
\notag \\ 
& \hspace{10pt} \times 
\tr \left[
(-2i\gamma_{5} \tau_{i}) \frac{1}{\cancel{k} - 2 \sigma_{0} - 4 t_{0} \gamma^{1} \gamma^{2} \tau_{3} }
\right.
\notag \\ 
& \hspace{20pt} \times
\left.
(-2i\gamma_{5} \tau_{i}) \frac{1}{\cancel{k} + \cancel{p} - 2 \sigma_{0} - 4 t_{0} \gamma^{1} \gamma^{2} \tau_{3} }
\right],
\label{eq:tpvf_massless_pi}
\\ 
\Gamma^{\mu\nu, \rho \sigma}_{t_{i = 1, 2}} (p) 
&= 
- \frac{4}{G_{T}}
( g^{\mu\rho} g^{\nu\sigma} - g^{\mu \sigma} g^{\nu\rho} )
- \int \frac{\dd^{4} k}{i (2 \pi)^{4}}
\notag \\ 
& \hspace{10pt} \times 
\tr \left[
(-4 \gamma^{\mu} \gamma^{\nu} \tau_{i}) \frac{1}{\cancel{k} - 2 \sigma_{0} - 4 t_{0} \gamma^{1} \gamma^{2} \tau_{3} }
\right.
\notag \\ 
& \hspace{20pt} \times
\left.
(-4 \gamma^{\rho} \gamma^{\sigma} \tau_{i}) \frac{1}{\cancel{k} + \cancel{p} - 2 \sigma_{0} - 4 t_{0} \gamma^{1} \gamma^{2} \tau_{3} }
\right]. 
\label{eq:tpvf_massless_t}
\end{align}
With $\mu = \rho = 1$, $\nu = \sigma = 2$, and $p \to 0$, and following the same calculations as in Appendix \ref{sec:appendixB} , we obtain
\begin{align}
&\Gamma_{\pi_{i = 1, 2}} (p \rightarrow 0) 
\notag \\ 
=& - \frac{2}{G_{S}} - \int \frac{\dd^{4} k}{i (2 \pi)^{4}} \tr 
\left[
(-2i\gamma_{5} \tau_{i}) \frac{1}{\cancel{k} - 2 \sigma_{0} - 4 t_{0} \gamma^{1} \gamma^{2} \tau_{3} }
\right.
\notag \\ 
& \hspace{80pt} \times 
\left.
(-2i\gamma_{5} \tau_{i}) \frac{1}{\cancel{k} - 2 \sigma_{0} - 4 t_{0} \gamma^{1} \gamma^{2} \tau_{3} }
\right]
\notag \\ 
=& - \frac{2}{G_{S}} - 96 \int \frac{\dd^{4} k}{i (2 \pi)^{4}}
\notag \\ 
& \hspace{20pt} \times 
 \frac{ k^{2} - 4 \sigma_{0}^{2} - 16 t_{0}^{2} }{ (k^{2} - 4 \sigma_{0}^{2} + 16 t_{0}^{2})^{2} + 64 t_{0}^{2} ( k_{1}^{2} + k_{2} ^{2} + 4 \sigma_{0}^{2} ) }
 \notag \\ 
=& - \frac{2}{G_{S}} \left(1 + 48 G_{S} \int \frac{\dd^{4} k}{i (2 \pi)^{4}}
\right.
\notag \\ 
& \hspace{20pt} \times 
\left.
 \frac{ k^{2} - 4 \sigma_{0}^{2} + 16 t_{0}^{2} }{ (k^{2} - 4 \sigma_{0}^{2} - 16 t_{0}^{2})^{2} - 64 t_{0}^{2} ( k_{1}^{2} + k_{2} ^{2} + 4 \sigma_{0}^{2} ) } 
 \right)
\notag \\ 
=& 0,  
\label{eq:tpvf_p0_massless_pi}
\\ 
&\Gamma_{t_{i = 1, 2}}^{12, 12} (p \rightarrow 0) 
\notag \\ 
=& - \frac{4}{G_{T}} - \int \frac{\dd^{4} k}{i (2 \pi)^{4}} \tr 
\left[
(-4 \gamma^{1} \gamma^{2} \tau_{i}) \frac{1}{\cancel{k} - 2 \sigma_{0} - 4 t_{0} \gamma^{1} \gamma^{2} \tau_{3} }
\right.
\notag \\ 
& \hspace{80pt} \times 
\left.
(-4 \gamma^{1} \gamma^{2} \tau_{i}) \frac{1}{\cancel{k} - 2 \sigma_{0} - 4 t_{0} \gamma^{1} \gamma^{2} \tau_{3} }
\right]
\notag \\ 
=& - \frac{4}{G_{T}} + 4 \cdot 96 \int \frac{\dd^{4} k}{i (2 \pi)^{4}}
\notag \\ 
& \hspace{20pt} \times 
\frac{( k^{2} - 4 \sigma_{0}^{2} - 16 t_{0}^{2} ) + 2 (k_{1}^{2} + k_{2}^{2} + 4 \sigma_{0}^{2} ) }{ (k^{2} - 4 \sigma_{0}^{2} - 16 t_{0}^{2})^{2} - 64 t_{0}^{2} ( k_{1}^{2} + k_{2} ^{2} + 4 \sigma_{0}^{2} ) }
\notag \\ 
=& - \frac{4}{G_{T}} \left( 1- 96 G_{T} \int \frac{\dd^{4} k}{i (2 \pi)^{4}}
\right.
\notag \\ 
& \hspace{20pt} \times 
\left.
\frac{( k^{2} - 4 \sigma_{0}^{2} - 16 t_{0}^{2} ) + 2 (k_{1}^{2} + k_{2}^{2} + 4 \sigma_{0}^{2} ) }{ (k^{2} - 4 \sigma_{0}^{2} - 16 t_{0}^{2})^{2} - 64 t_{0}^{2} ( k_{1}^{2} + k_{2} ^{2} + 4 \sigma_{0}^{2} ) } \right)
\notag \\ 
=& 0. 
\label{eq:tpvf_p0_massless_t}
\end{align}
Here, in the final equality of \eqref{eq:tpvf_p0_massless_pi} and \eqref{eq:tpvf_p0_massless_t}, we used the gap equations \eqref{eq:gap_chiral_coexist} and \eqref{eq:gap_tensor_coexist}, respectively.
Therefore, from \eqref{eq:tpvf_and_mass}, we find that the masses of $\pi_{i = 1, 2}$ and $t_{i = 1, 2}^{12}$ are zero, confirming that they are massless modes. 
Thus, we will consider the other two-point vertex functions from now on.
Similar to \eqref{eq:tpvf_massless_pi} and \eqref{eq:tpvf_massless_t}, we obtain the following expressions:
\begin{align}
\Gamma_{\pi_{3}} (p) 
&= - \frac{2}{G_{S}} - \int \frac{\dd^{4} k}{i (2 \pi)^{4}} 
\notag \\ 
& \hspace{10pt} \times 
\tr \left[
(-2i\gamma_{5} \tau_{3}) \frac{1}{\cancel{k} - 2 \sigma_{0} - 4 i t_{0} \gamma^{1} \gamma^{2} \tau_{3} }
\right.
\notag \\ 
& \hspace{20pt} \times
\left.
(-2i\gamma_{5} \tau_{3}) \frac{1}{\cancel{k} + \cancel{p} - 2 \sigma_{0} - 4 i t_{0} \gamma^{1} \gamma^{2} \tau_{3} }
\right], 
\label{eq:tpvf_pi3-pi3}
\end{align}
\begin{align}
\Gamma^{\mu\nu}_{\pi_{3} \varpi} (p)
&= - \int \frac{\dd^{4} k}{i (2 \pi)^{4}} 
\notag \\ 
& \hspace{10pt} \times 
\tr \left[
(-4i\gamma_{5} \gamma^{\mu} \gamma^{\nu}) \frac{1}{\cancel{k} - 2 \sigma_{0} - 4 i t_{0} \gamma^{1} \gamma^{2} \tau_{3} }
\right.
\notag \\ 
& \hspace{20pt} \times 
\left.
(-2i\gamma_{5} \tau_{3}) \frac{1}{\cancel{k} + \cancel{p} - 2 \sigma_{0} - 4 i t_{0} \gamma^{1} \gamma^{2} \tau_{3} }
\right], 
\label{eq:tpvf_pi3-varpi}
\\ 
\Gamma^{\mu\nu}_{\varpi \pi_{3}} (p) 
&= - \int \frac{\dd^{4} k}{i (2 \pi)^{4}}  
\notag \\ 
& \hspace{10pt} \times 
\tr \left[
(-2i\gamma_{5} \tau_{3}) \frac{1}{\cancel{k} - 2 \sigma_{0} - 4 i t_{0} \gamma^{1} \gamma^{2} \tau_{3} }
\right.
\notag \\ 
& \hspace{20pt} \times 
\left.
(-4i\gamma_{5} \gamma^{\mu} \gamma^{\nu}) \frac{1}{\cancel{k} + \cancel{p} - 2 \sigma_{0} - 4 i t_{0} \gamma^{1} \gamma^{2} \tau_{3} }
\right], 
\label{eq:tpvf_varpi-pi3}
\\ 
\Gamma^{\mu\nu, \rho\sigma}_{ \varpi } (p) 
&= - \frac{4}{G_{T}} ( g^{\mu\rho} g^{\nu\sigma} - g^{\mu\sigma} g^{\nu\rho} )
- \int \frac{\dd^{4} k}{i (2 \pi)^{4}} 
\notag \\ 
& \hspace{10pt} \times
\tr \left[
(-4i\gamma_{5} \gamma^{\mu} \gamma^{\nu}) \frac{1}{\cancel{k} - 2 \sigma_{0} - 4 i t_{0} \gamma^{1} \gamma^{2} \tau_{3} }
\right.
\notag \\ 
& \hspace{20pt} \times 
\left.
(-4i\gamma_{5} \gamma^{\rho} \gamma^{\sigma}) \frac{1}{\cancel{k} + \cancel{p} - 2 \sigma_{0} - 4 i t_{0} \gamma^{1} \gamma^{2} \tau_{3} }
\right], 
\label{eq:tpvf_varpi-varpi}
\end{align}
where $\cancel{k} = \gamma^{\mu} k_{\mu}$ and so on. 
It should be necessary to consider the mode mixing between $\pi_3$ and $\varpi_{12}$.

By performing the calculations provided in the Appendix \ref{sec:appendixB}, taking the limit $p \to 0$ in \eqref{eq:tpvf_pi3-pi3}-\eqref{eq:tpvf_varpi-varpi}, and using the gap equations \eqref{eq:gap_chiral_coexist} and \eqref{eq:gap_tensor_coexist}, we obtain the following expressions: 
\begin{align}
&\Gamma_{\pi_{3}} (p \to 0) 
\notag \\ 
=& 
- \frac{2}{G_{S}} - \int \frac{\dd^{4} k}{i (2 \pi)^{4}} \tr 
\left[
(-2i\gamma_{5} \tau_{3}) \frac{1}{\cancel{k} - 2 \sigma_{0} - 4 i t_{0} \gamma^{1} \gamma^{2} \tau_{3} }
\right.
\notag \\ 
& \hspace{80pt} \times 
\left.
(-2i\gamma_{5} \tau_{3}) \frac{1}{\cancel{k} - 2 \sigma_{0} - 4 i t_{0} \gamma^{1} \gamma^{2} \tau_{3} }
\right]
\notag \\ 
=&
96 \int \frac{\dd^{4} k}{i (2 \pi)^{4}} 
 \frac{32 t_{0}^{2} }{ (k^{2} - 4 \sigma_{0}^{2} - 16 t_{0}^{2})^{2} - 64t_{0}^{2} (k_{1}^{2} + k_{2}^{2} + 4 \sigma_{0}^{2}) }, 
\label{eq:tpvf_p0_pi3}
\end{align}
\begin{align}
&\Gamma^{12}_{\pi_{3} \varpi} (p \to 0)
\notag \\ 
=& 
- \int \frac{\dd^{4} k}{i (2 \pi)^{4}} \tr 
\left[
(-4i\gamma_{5} \gamma^{1} \gamma^{2}) \frac{1}{\cancel{k} - 2 \sigma_{0} - 4 i t_{0} \gamma^{1} \gamma^{2} \tau_{3} }
\right.
\notag \\ 
& \hspace{80pt} \times 
\left.
(-2i\gamma_{5} \tau_{3}) \frac{1}{\cancel{k} - 2 \sigma_{0} - 4 i t_{0} \gamma^{1} \gamma^{2} \tau_{3} }
\right]
\notag \\ 
=&
96 \int \frac{\dd^{4} k}{i (2 \pi)^{4}} \frac{ 32 i \sigma_{0} t_{0} }{ (k^{2} - 4 \sigma_{0}^{2} - 16 t_{0}^{2})^{2} - 64t_{0}^{2} (k_{1}^{2} + k_{2}^{2} + 4 \sigma_{0}^{2}) }
\notag \\ 
=& 
\Gamma^{12}_{\varpi \pi_{3}} (p \to 0), 
\label{eq:tpvf_p0_pi3-varpi}
\end{align}
\begin{align}
&\Gamma^{12, 12}_{ \varpi } (p \to 0) 
\notag \\ 
=&
- \frac{4}{G_{T}}
- \int \frac{\dd^{4} k}{i (2 \pi)^{4}}
\left[
(-4i\gamma_{5} \gamma^{1} \gamma^{2}) \frac{1}{\cancel{k} - 2 \sigma_{0} - 4 i t_{0} \gamma^{1} \gamma^{2} \tau_{3} }
\right.
\notag \\ 
& \hspace{80pt} \times 
\left.
(-4i\gamma_{5} \gamma^{1} \gamma^{2}) \frac{1}{\cancel{k} - 2 \sigma_{0} - 4 i t_{0} \gamma^{1} \gamma^{2} \tau_{3} }
\right]
\nonumber \\ 
=& 
96 \int \frac{\dd^{4} k}{i (2 \pi)^{4}}
\frac{ - 32 \sigma_{0}^{2} }{ (k^{2} - 4 \sigma_{0}^{2} - 16 t_{0}^{2})^{2} - 64t_{0}^{2} (k_{1}^{2} + k_{2}^{2} + 4 \sigma_{0}^{2}) }. 
\label{eq:tpvf_p0_varpi}
\end{align}
Thus, the mass term $^{t}\bm{x} M^{2}_{\pi_{3} \varpi^{12}} \bm{x}$ can be considered through the following equation derived from the two-point vertex functions:
\begin{align}
M^{2}_{\pi_{3} \varpi^{12}}
&=
   \begin{pmatrix}
   \Gamma_{\pi_{3} \pi_{3}} (p \longrightarrow 0) & \Gamma^{12}_{\pi_{3} \varpi } (p \longrightarrow 0) \\[4pt]
   \Gamma^{12}_{\varpi \pi_{3}} (p \longrightarrow 0)  & \Gamma^{12, 12}_{ \varpi } (p \longrightarrow 0) 
   \end{pmatrix}
\nonumber \\ 
&= 32 K
   \begin{pmatrix}
    t_{0}^{2} &  i \sigma_{0} t_{0}          \\[4pt]
    i \sigma_{0} t_{0}  & - \sigma_{0}^{2}
   \end{pmatrix},
\label{eq:mass_matrix_non-Hermitian}
\end{align}
where
\begin{align}
K 
&= 
96 \int \frac{\dd^{4} k}{i (2 \pi)^{4}}
\frac{ 1 }{ (k^{2} - 4 \sigma_{0}^{2} - 16 t_{0}^{2})^{2} - 64t_{0}^{2} (k_{1}^{2} + k_{2}^{2} + 4 \sigma_{0}^{2}) }
\label{eq:mass_matrix_integral}
\end{align}
and 
\begin{align}
	\bm{x} = \begin{pmatrix}
			\pi_{3} \\ 
			\varpi^{12}
			\end{pmatrix}. 
\label{eq:state_pi3_p12}
\end{align}
In the case we are considering where $\mu \neq \nu$, we have
\begin{align}
( \varpi^{\mu \nu} )^{\dagger} = -  \varpi^{\mu \nu}. 
\label{eq:p-meson_spacecomponent}
\end{align}
Therefore, to obtain an Hermitian field, we redefine $\varpi^{12}$ as
\begin{align}
\varpi^{12} = i \varpi^{12}_{h}. 
\label{eq:p-meson_redefine}
\end{align}
Thus, transforming the state $\bm{x}$ in \eqref{eq:state_pi3_p12} into a real-valued state $\bm{x}'$, we obtain the mass matrix $\mu^{2}$ as follows:
\begin{align}
^{t}\bm{x} M_{\pi_{3} \varpi^{12}}^{2} \bm{x} 
&= 
\begin{pmatrix}
\pi_3 & \varpi^{12}_{h}
\end{pmatrix}
\begin{pmatrix}
1 & 0 \\ 
0 & i 
\end{pmatrix}
M_{\pi_{3} p^{12}}^{2}
\begin{pmatrix}
1 & 0 \\ 
0 & i 
\end{pmatrix}
\begin{pmatrix}
\pi_3 \\ 
\varpi^{12}_{h}
\end{pmatrix}
\notag \\ 
&= 
32K
\begin{pmatrix}
\pi_3 & \varpi^{12}_{h}
\end{pmatrix}
\begin{pmatrix}
t_0^{2} & - \sigma_{0} t_{0} \\ 
- \sigma_{0} t_{0} & \sigma_{0}^{2}
\end{pmatrix}
\begin{pmatrix}
\pi_3 \\ 
\varpi^{12}_{h}
\end{pmatrix}
\notag \\ 
&=
\,^{t}\bm{x}' \mu^{2} \bm{x}'. 
\label{eq:mass_term}
\end{align}
Here, 
\begin{align}
\mu^{2} =
	32K
	\begin{pmatrix}
	t_0^{2} & - \sigma_{0} t_{0} \\ 
	- \sigma_{0} t_{0} & \sigma_{0}^{2}
	\end{pmatrix}
\label{eq:mass_matrix_Hermitian}
\end{align}
and
\begin{align}
\bm{x}' =
	\begin{pmatrix}
	\pi_3 \\ 
	\varpi^{12}_{h}
	\end{pmatrix}. 
\label{eq:state_real-valued}
\end{align}

In this case, the mass matrix $\mu^{2}$ is Hermitian. 
Therefore, it can be diagonalized by a unitary matrix $U$ as follows:
\begin{align}
   \begin{split}
U
&= 
\frac{1}{ \sqrt{t_{0}^{2} + \sigma_{0}^{2}} }
\begin{pmatrix}
- \sigma_{0} & - t_{0} \\ 
- t_{0} & \sigma_{0}
\end{pmatrix}, 
\\ 
U^{-1}
&=
U^{\dagger}
= 
\frac{1}{ \sqrt{t_{0}^{2} + \sigma_{0}^{2}} }
\begin{pmatrix}
- \sigma_{0} & - t_{0} \\ 
- t_{0} & \sigma_{0}
\end{pmatrix}, 
\\ 
U^{-1} \mu^{2} U 
&= 
\begin{pmatrix}
0 & 0 \\ 
0 & 32K (t_{0}^{2} + \sigma_{0}^{2})
\end{pmatrix},
\\
U^{-1} \bm{x'}
&=\frac{1}{\sqrt{t_0^2+\sigma_0^2}} 
\begin{pmatrix}
-\sigma_0\pi_{3}-t_0 \varpi^{12}_{h} \\ 
-t_0\pi_3+\sigma_0 \varpi^{12}_{h}
\end{pmatrix}.
   \end{split}
\label{eq:mass_matrix_diag}
\end{align}
Thus, we conclude that $- ( \sigma_{0} \pi_{3} + t_{0} \varpi^{12}_{h} ) / \sqrt{t_{0}^{2} + \sigma_{0}^{2}}$ is the massless mode and $(-t_{0} \pi_{3} + \sigma_{0} \varpi^{12}_{h} ) / \sqrt{t_{0}^{2} + \sigma_{0}^{2}}$ is the massive mode with mass $\sqrt{32K (t_{0}^{2} + \sigma_{0}^{2})}$. 
Therefore, the NG modes are $\pi_1, \pi_2, t_{1}^{12}, t_2^{12}$, and $-(\sigma_0\pi_3+t_0 \varpi^{12}_{h} )/\sqrt{t_0^2+\sigma_0^2}$ in the case that the chiral condensate $\sigma_0$ and the tensor (spin polarized) condensate $t_0$ coexist. 

%%%%%%%%%%%%%%%%%%%%%%%%%%%%%%%%%%%%%
%%%%%%%%%%%%%%%%%%%%%%%%%%%%%%%%%%%%%

\section{\label{sec:level3}Sponteneous breaking of chiral symmetry and space symmetry}

In Sec. \ref{sec:level2}, the NG modes are investigated by means of the two-point vertex functions by searching the zero points of the inverse propagators.
Hereafter, in Secs. \ref{sec:level3} and \ref{sec:level4}, we find the NG modes by the algebraic method.
In Sec. \ref{sec:level3-1}, the NG modes developed in Sec. \ref{sec:level2} are reproduced in the case that the chiral and the tensor condensates exist. 
In Sec. \ref{sec:level3-2}, the NG modes are investigated in the case of the condensate with space dependence, namely the inhomogeneous chiral condensate.

\subsection{\label{sec:level3-1}The case of the chiral and space symmetry breaking by the homogeneous chiral condensate and the tensor condensate}

We consider the chiral $ su_{A}(2) \times su_{V}(2)$ transformations:
\begin{align}
\psi 
\longrightarrow \ & e^{i \theta_{a} \gamma_{5}\tau^{a}/2} \psi 
\approx \  \psi + \theta_{a} \delta^{a}_{A} \psi,
& \notag \\ 
& \hspace{80pt} su_{A}(2) \hspace{12pt} \text{transformation,}
\label{eq:su_A(2)trans}
\\ 
\psi 
\longrightarrow \ & e^{i \phi_{a} \tau^{a}/2} \psi 
\approx \  \psi + \phi_{a} \delta^{a}_{V} \psi,
& \notag \\ 
& \hspace{80pt} su_{V}(2) \hspace{12pt} \text{transformation.}
\label{eq:su_V(2)trans}
\end{align}
Here, $\delta^{a}_{A} \psi = i \gamma_{5} (\tau^{a}/2) \psi$ and $\delta^{a}_{V} \psi = i (\tau^{a}/2) \psi$, where $a$ represents the isospin indices. 
From \eqref{eq:su_A(2)trans} and \eqref{eq:su_V(2)trans}, the Noether current $j_{a}^{\mu}$ and the Noether charge $Q^{a}$, which arise due to the symmetry under these transformations, are given by
\begin{align}
     \begin{split}
j_{A, a}^{\mu} 
&= 
\frac{\partial \mathscr{L}_{ \mathrm{NJL, T} } }{\partial (\partial_{\mu} \psi)} \delta^{a}_{A} \psi 
= 
- \bar{\psi} \gamma^{\mu} \gamma_{5} \frac{\tau^{a}}{2} \psi , 
\\ 
j_{V, a}^{\mu} 
&= 
\frac{\partial \mathscr{L}_{ \mathrm{NJL, T} } }{\partial (\partial_{\mu} \psi)} \delta^{a}_{V} \psi 
= 
- \bar{\psi} \gamma^{\mu} \frac{\tau^{a}}{2} \psi , 
     \end{split}
\label{eq:NoetherCurrent_chiral_su(2)}
\\ 
     \begin{split}
Q^{a}_{A} 
&=
\int \dd^{3} x \, j^{\mu = 0}_{A, a} 
= - \int \dd^{3} x \, \bar{\psi} \gamma^{0} \gamma_{5} \frac{\tau^{a}}{2} \psi,  
\\ 
Q^{a}_{V} 
&=
\int \dd^{3} x \, j^{\mu = 0}_{V, a} 
= - \int \dd^{3} x \, \bar{\psi} \gamma^{0} \frac{\tau^{a}}{2} \psi. 
     \end{split}
\label{eq:NoetherCharge_chiral_su(2)}
\end{align}
Now, under the chiral transformations, the following commutation relations can be derived:
\begin{align}
     \begin{split}
[i Q_{A}^{a}, \bar{\psi} i \gamma_{5} \tau^{b} \psi] 
&= - \delta^{ab} \bar{\psi} \psi, 
\\ 
[i Q_{A}^{a}, \bar{\psi} \psi] 
&= \bar{\psi} i \gamma_{5} \tau^{a} \psi, 
\\ 
[i Q_{V}^{a}, \bar{\psi} i \gamma_{5} \tau^{b} \psi] 
&= \tensor{\epsilon}{^a^b^c} \bar{\psi} i \gamma_{5} \tau^{c} \psi, 
\\ 
[i Q_{V}^{a}, \bar{\psi} \psi] 
&= 0. 
     \end{split}
\label{eq:comm_relation_scalar}
\end{align}
If the quark-antiquark chiral condensate realizes a nonzero value, namely $\langle \bar{\psi} \psi \rangle \neq 0$, we obtain the following from the first equation of \eqref{eq:comm_relation_scalar}:
\begin{align}
\langle [i Q_{A}^{a}, \bar{\psi} i \gamma_{5} \tau^{b} \psi]  \rangle =  - \delta^{ab} \langle \bar{\psi} \psi \rangle \neq 0, 
\label{eq:SSB_chiral_pseudoscalar}
\end{align}
where $\langle\cdots\rangle$ means the expectation value of the field. 
Here, $\delta^{ab}$ represents the Kronecker delta. 
Therefore, the chiral $su_{A} (2)$-symmetry with the generators $Q_{A}^{a} (a = 1, 2, 3)$ is spontaneously broken, and the NG modes are
\begin{align}
\bar{\psi} i \gamma_{5} \tau^{b} \psi \approx \pi^{b}, 
\label{eq:NG_pseudoscalar} 
\end{align}
where $\bar{\psi} i \gamma_{5} \tau^{b} \psi$ for each $b = 1, 2, 3$ corresponds to the NG boson, namely pion.

Similarly, considering tensor condensate, the following commutation relations hold:
\begin{align}
     \begin{split}
[i Q_{A}^{a}, \bar{\psi} \gamma^{\mu} \gamma^{\nu} \tau^{b} \psi] 
&= \delta^{ab} i \bar{\psi} \gamma_{5} \gamma^{\mu} \gamma^{\nu} \psi, 
\\ 
[i Q_{A}^{a}, \bar{\psi} i \gamma_{5} \gamma^{\mu} \gamma^{\nu} \psi] 
&= - \bar{\psi} \gamma^{\mu} \gamma^{\nu} \tau^{a} \psi, 
\\ 
[i Q_{V}^{a}, \bar{\psi} \gamma^{\mu} \gamma^{\nu} \tau^{b} \psi] 
&= \tensor{\epsilon}{^a^b^c} \bar{\psi} \gamma^{\mu} \gamma^{\nu} \tau^{c} \psi, 
\\ 
[i Q_{V}^{a}, \bar{\psi} i \gamma_{5} \gamma^{\mu} \gamma^{\nu} \psi] 
&= 0. 
     \end{split}
\label{eq:comm_relation_tensor}
\end{align}
Here, $\tensor{\epsilon}{^a^b^c}$ is the completely antisymmetric tensor. 
Therefore, since the tensor-type quark-antiquark condensate realizes a nonzero value, namely $\langle i \bar{\psi} \gamma^{1} \gamma^{2} \tau^{3} \psi \rangle \neq 0$, we obtain the following from the second and third equations of \eqref{eq:comm_relation_tensor}:
\begin{align}
     \begin{split}
\langle [i Q_{A}^{a = 3}, \bar{\psi} \gamma_{5} \gamma^{1} \gamma^{2} \psi]  \rangle 
&= \langle i \bar{\psi} \gamma^{1} \gamma^{2} \tau^{3} \psi \rangle 
\neq 0, 
\\ 
\langle [i Q_{V}^{a = 1}, - i \bar{\psi} \gamma^{1} \gamma^{2} \tau^{b = 2} \psi]  \rangle 
&=  - \langle i \bar{\psi} \gamma^{1} \gamma^{2} \tau^{3} \psi \rangle 
\neq 0, 
\\ 
\langle [i Q_{V}^{a = 2}, -i \bar{\psi} \gamma^{1} \gamma^{2} \tau^{b = 1} \psi]  \rangle 
&=  \langle i \bar{\psi} \gamma^{1} \gamma^{2} \tau^{3} \psi \rangle 
\neq 0. 
     \end{split}
\label{eq:SSB_chiral_tensor}
\end{align}
Therefore, from the second and third equations of \eqref{eq:SSB_chiral_tensor}, the chiral $su_{V}(2)$-symmetry with generators $Q_{V}^{a = 1}$ and $Q_{V}^{a = 2}$ is spontaneously broken, and the corresponding NG modes
\begin{align}
	\begin{split}
\bar{\psi} \gamma^{1} \gamma^{2} \tau^{2} \psi \approx t^{12}_{2}
\\ 
\bar{\psi} \gamma^{1} \gamma^{2} \tau^{1} \psi \approx t^{12}_{1}
	\end{split}
	\label{eq:NG_tensor}
\end{align}
appear. 
Similarly, the chiral $su_{A}(2)$-symmetry with the generator $Q_{A}^{a=3}$ is spontaneously broken, and the corresponding NG mode 
\begin{align}
\bar{\psi} \gamma_{5} \gamma^{1} \gamma^{2} \psi \approx \varpi^{12}_{h}
\label{eq:NG_pseudotensor}
\end{align}
appears. 
Now, from
\begin{align}
\varpi^{12} &= - G_{T} \bar{\psi} i \gamma_{5} \gamma^{1} \gamma^{2} \psi = i \varpi^{12}_{h}, 
\label{eq:pseudotensor_field}
\end{align}
since $\varpi^{12}$ is purely imaginary, we redefine it as a real field $\varpi^{12}_{h}$:  
\begin{align}
\varpi^{12}_{h} &= - G_{T} \bar{\psi} \gamma_{5} \gamma^{1} \gamma^{2} \psi. 
\label{eq:pseudotensor_field_redef}
\end{align}

It seems that the generator $Q_{A}^{a = 3}$ causes two NG modes, namely $\pi^{3}$ and $\varpi^{12}_{h}$. 
This suggests that the $\pi^{3}-\varpi^{12}_{h}$ mixing occurs. 
In fact, from \eqref{eq:SSB_chiral_pseudoscalar} and the first equations of \eqref{eq:SSB_chiral_tensor}, 
\begin{align}
     \begin{split}
\langle [i Q_{A}^{a = 3}, - G_{S} \bar{\psi} i \gamma_{5} \tau^{3} \psi]  \rangle 
&=
\langle [i Q_{A}^{a = 3}, \pi^{3}]  \rangle 
\\ 
&=  
G_{S} \langle \bar{\psi} \psi \rangle 
= 
- \sigma_{0}, 
\\ 
\langle [i Q_{A}^{a = 3}, - G_{T} \bar{\psi} \gamma_{5} \gamma^{1} \gamma^{2} \psi]  \rangle
&=
\langle [i Q_{A}^{a = 3}, \varpi^{12}_{h}]  \rangle
\\ 
&= - G_{T} \langle i \bar{\psi} \gamma^{1} \gamma^{2} \tau^{3} \psi \rangle 
= - t_{0}, 
     \end{split}
\label{eq:SSB_QA3}
\end{align}
we obtain the following commutation relations: 
\begin{align}
     \begin{split}
\langle [i Q_{A}^{a = 3}, - \frac{\sigma_{0} \pi^{3} + t_{0} \varpi^{12}_{h} }{\sqrt{\sigma_{0}^{2} + t_{0}^{2} } }]  \rangle 
&=
\frac{\sigma_{0}^{2} + t_{0}^{2}}{\sqrt{\sigma_{0}^{2} + t_{0}^{2}}}
\neq 0, 
\\ 
\langle [i Q_{A}^{a = 3}, \frac{- t_{0} \pi^{3} + \sigma_{0} \varpi^{12}_{h} }{\sqrt{\sigma_{0}^{2} + t_{0}^{2} } }]  \rangle 
&=
\frac{t_{0}\sigma_{0} - \sigma_{0} t_{0} }{\sqrt{\sigma_{0}^{2} + t_{0}^{2}}}
= 
0. 
     \end{split}
     \label{eq:NG_mixing}
\end{align}
This indicates that $-(\sigma_{0} \pi^{3} + t_{0} \varpi^{12}_{h})/\sqrt{\sigma_{0}^{2} + t_{0}^{2} }$ becomes a NG mode and $(-t_{0} \pi^{3} + \sigma_{0} \varpi^{12}_{h})/\sqrt{\sigma_{0}^{2} + t_{0}^{2} }$ becomes the massive mode due to $\pi^{3}$-$\varpi^{12}_{h}$ mixing. 
Therefore, from \eqref{eq:NG_pseudoscalar}, \eqref{eq:NG_tensor}, and \eqref{eq:NG_mixing}, we find that the following five NG modes correspond to the breaking of chiral symmetry: 
\begin{align}
\pi^{1}, \pi^{2}, t^{12}_{1}, t^{12}_{2},  - \frac{\sigma_{0} \pi^{3} + t_{0} \varpi^{12}_{h} }{\sqrt{\sigma_{0}^{2} + t_{0}^{2} } }. 
\label{eq:NG_chiral}
\end{align}
This result is consistent with that obtained in Sec. \ref{sec:level2}. 

%%%%%%%%%%%%%%%%%%%%%%%%%%%%%%%%%%%%%
%%%%%%%%%%%%%%%%%%%%%%%%%%%%%%%%%%%%%

Next, under the Lagrangian \eqref{eq:L_NJL_T}, we consider spacetime translations and the following Lorentz transformations:
\begin{align}
x^{\rho} 
&\longrightarrow& 
{x'}^{\rho} 
&= x^{\rho} + \tensor{\epsilon}{^{\rho}^{\sigma}} x_{\sigma}
\label{Lorentz_transformation}
\\ 
\psi_{i} (x) 
&\longrightarrow& 
\psi_{i}'(x')  &=
1 - \frac{i}{2} \tensor{\epsilon}{^{\rho}^{\sigma}} \left[
\tensor{\left( \tensor{S}{_{\rho}_{\sigma}}\right)}{_{i}^{j}} +
\tensor{L}{_{\rho}_{\sigma}} \tensor{\delta}{_{i}^{j}} 
\right]
\psi_{j} (x). 
\label{eq:field-trans_Lorentz}
\end{align}
Here
\begin{align}
	\begin{split}
\tensor{L}{_{\rho}_{\sigma}} 
&= i \left(  x_{\rho} \partial_{\sigma} - x_{\sigma} \partial_{\rho} \right), 
\\ 
\tensor{S}{_{\rho}_{\sigma}}
&= \frac{i}{4} [\gamma_{\rho}, \gamma_{\sigma}] . 
	\end{split}
\label{eq:factor_Lorentz}
\end{align}
The Noether currents for spacetime translations and Lorentz transformations are given by
\begin{align}
\tensor{T}{_{\rho}^{\mu}} (x)
&= 
\frac{\partial \mathscr{L}_{ \mathrm{NJL, T} } }{ \partial (\partial_{\mu} \psi_{i} (x) )} \partial_{\rho} \psi_{i} - \delta^{\mu}_{\rho} \mathscr{L}_{ \mathrm{NJL, T} }, 
\label{eq:Noether_current_translation}
\\ 
\tensor{\mathscr{M}}{_{\rho}_{\sigma}^{\mu}} (x)
&= 
\frac{\partial \mathscr{L}_{ \mathrm{NJL, T} } }{ \partial (\partial_{\mu} \psi_{i} )} 
\left( 
- i \tensor{\left( \tensor{S}{_{\rho}_{\sigma}} \right)}{_{i}^{j}} \psi_{j}
+ \left(  x_{\rho} \partial_{\sigma} - x_{\sigma} \partial_{\rho} \right)  \psi_{i}
\right)
\notag \\ 
& \hspace{10pt}
- (x_{\rho} \delta^{\mu}_{\sigma} - x_{\sigma} \delta^{\mu}_{\rho} ) \mathscr{L}_{ \mathrm{NJL, T} }
\notag \\ 
&=
x_{\rho} \tensor{T}{_{\sigma}^{\mu}} - x_{\sigma} \tensor{T}{_{\rho}^{\mu}} - i 
\frac{\partial \mathscr{L}_{ \mathrm{NJL, T} } }{ \partial (\partial_{\mu} \psi_{i} )} \tensor{\left( \tensor{S}{_{\rho}_{\sigma}} \right)}{_{i}^{j}} \psi_{j}. 
\label{eq:Noether_current_Lorentz}
\end{align}
Therefore, the translation generator $P_{\rho}$ and the Lorentz transformation generator $M_{\rho \sigma}$ can be obtained from the Noether currents \eqref{eq:Noether_current_translation} and \eqref{eq:Noether_current_Lorentz} as follows:
\begin{align}
P_{\rho}
&=
\int \dd^{3} x \, 
\left( \bar{\psi} i \gamma^{0} \partial_{\rho} \psi - \delta^{0}_{\rho} \mathscr{L}_{ \mathrm{NJL, T} }
\right), 
\label{eq:translation_generator}
\\ 
M_{\rho \sigma}
&=
\int \dd^{3} x \, 
( 
x_{\rho} \bar{\psi} i \gamma^{0} \partial_{\sigma} \psi - x_{\sigma} \bar{\psi} i \gamma^{0} \partial_{\rho} \psi 
\notag \\ 
&\hspace{10pt}
- i \cdot  
\bar{\psi}^{i} i \gamma^{0} \tensor{\left( \tensor{S}{_{\rho}_{\sigma}} \right)}{_{i}^{j}} \psi_{j}
- 
(x_{\rho} \delta^{0}_{\sigma} - x_{\sigma}  \delta^{0}_{\rho}) \mathscr{L}_{ \mathrm{NJL, T} }
). 
\label{eq:Lorentz_generator}
\end{align}
Here, we consider only spatial rotations, excluding time translations and boosts, thus setting $\rho = a, \sigma = b, a \neq b (a, b = 1, 2, 3) $. In this case,
\begin{align}
S_{ab}
= \frac{i}{2} \gamma_{a} \gamma_{b}
\label{eq:factor_Lorentz_spatial}
\end{align}
can be used.
In this context, from \eqref{eq:translation_generator} and \eqref{eq:Lorentz_generator}, the generators of spatial translations $P_{a}$ and spatial rotations $M_{ab}$ are given by
\begin{align}
P_{a}
&=
\int \dd^{3} x \, \bar{\psi} i \gamma^{0} \partial_{a} \psi , 
\label{eq:tlans_generator}
\\ 
M_{ab}
&=
\int \dd^{3} x \, 
( 
x_{a} \bar{\psi} i \gamma^{0} \partial_{b} \psi 
- x_{b} \bar{\psi} i \gamma^{0} \partial_{a} \psi 
\notag \\ & \hspace{70pt}  
- i \cdot  
\bar{\psi}^{i} i \gamma^{0} \tensor{\left( \tensor{S}{_{a}_{b}} \right)}{_{i}^{j}} \psi_{j}
). 
\label{eq:rot_generator}
\end{align}
Then, the commutation relations between the generators of spatial translations $P_{a}$ and spatial rotations $M_{ab}$, and the fields $\bar{\psi} \gamma^{\mu} \gamma^{\nu} \tau^{b} \psi$ and $\bar{\psi} i \gamma_{5} \gamma^{\mu} \gamma^{\nu} \psi$ are obtained as follows:
\begin{align}
	\begin{split}
[iP_{a}, \bar{\psi} \gamma^{\mu} \gamma^{\nu} \tau^{b} \psi]
&= \partial_{a} (\bar{\psi} \gamma^{\mu} \gamma^{\nu} \tau^{b} \psi ), 
\\ 
[iP_{a}, \bar{\psi} i \gamma_{5} \gamma^{\mu} \gamma^{\nu} \psi]
&= \partial_{a} (\bar{\psi} i \gamma_{5} \gamma^{\mu} \gamma^{\nu} \psi ), 
\\ 
[iM_{ab}, \bar{\psi} \gamma^{\mu} \gamma^{\nu} \tau^{c} \psi]
&=
\partial_{b} ( x_{a} \bar{\psi} \gamma^{\mu} \gamma^{\nu} \tau^{c} \psi ) 
\\ & \hspace{10pt}  
- \partial_{a} ( x_{b} \bar{\psi} \gamma^{\mu} \gamma^{\nu} \tau^{c} \psi ) 
\\ & \hspace{20pt} 
+ \bar{\psi} [\tensor{S}{_{a}_{b}}, \gamma^{\mu} \gamma^{\nu} \tau^{c} ] \psi, 
\\ 
[iM_{ab}, \bar{\psi} i \gamma_{5} \gamma^{\mu} \gamma^{\nu} \psi]
&=
\partial_{b} ( x_{a} \bar{\psi} i \gamma_{5} \gamma^{\mu} \gamma^{\nu} \psi ) 
\\ & \hspace{10pt}
- \partial_{a} ( x_{b} \bar{\psi} i \gamma_{5} \gamma^{\mu} \gamma^{\nu} \psi ) 
\\ & \hspace{20pt}
+ \bar{\psi} [\tensor{S}{_{a}_{b}}, i \gamma_{5} \gamma^{\mu} \gamma^{\nu} ] \psi. 
	\end{split}
\label{eq:comm_space_tensor} 
\end{align} 
Then, if the tensor-type quark-antiquark condensate realizes a nonzero value, namely $\langle i \bar{\psi} \gamma^{1} \gamma^{2} \tau^{3} \psi \rangle = \langle \bar{\psi} \Sigma_{3} \tau_{3} \psi \rangle  = {\rm const}$, this situation corresponds to the quark spin polarization occurring in the direction of the third spatial component. 
Consequently, it is expected that the rotational symmetry involving the third spatial component, specifically the spatial rotations $(\rho, \sigma) = (3, 1), (2, 3)$, is broken. 
Therefore, we obtain the following equations:
\begin{align}
	\begin{split}
\langle [iP_{a}, \bar{\psi} \gamma^{\mu} \gamma^{\nu} \tau^{b} \psi] \rangle
&= \partial_{a} \langle \bar{\psi} \gamma^{\mu} \gamma^{\nu} \tau^{b} \psi \rangle
= 0, 
\\ 
\langle [iP_{a}, \bar{\psi} i \gamma_{5} \gamma^{\mu} \gamma^{\nu} \psi] \rangle
&= 0, 
\\ 
\langle [iM_{ab}, \bar{\psi} \gamma^{\mu} \gamma^{\nu} \tau^{c} \psi] \rangle
&=
\langle \bar{\psi} [\tensor{S}{_{a}_{b}}, \gamma^{\mu} \gamma^{\nu} \tau^{c} ] \psi \rangle,
\\ 
\langle [iM_{ab}, \bar{\psi} i \gamma_{5} \gamma^{\mu} \gamma^{\nu} \psi] \rangle
&=
\langle \bar{\psi} [\tensor{S}{_{a}_{b}}, i \gamma_{5} \gamma^{\mu} \gamma^{\nu} ] \psi \rangle.
	\end{split}
\label{eq:VEV_comm_rot_tensor} 
\end{align}
Therefore, while the spatial translation symmetry is not broken by the tensor condensate, the spatial rotational symmetry can be broken. 
In this context, noting that from \eqref{eq:factor_Lorentz_spatial}, the third equation of \eqref{eq:VEV_comm_rot_tensor} can be rewritten as follows:
\begin{align}
\langle \bar{\psi} [\tensor{S}{_{a}_{b}}, \gamma^{\mu} \gamma^{\nu} \tau^{c} ] \psi \rangle 
&=
i \langle \bar{\psi} 
[
- g_{a \alpha} \delta^{\nu}_{b} \gamma^{\alpha} \gamma^{\mu}
+ g_{a \alpha} \delta^{\mu}_{b} \gamma^{\alpha} \gamma^{\nu}
\notag \\ & \hspace{10pt}
- g_{b \beta} \delta^{\nu}_{a} \gamma^{\mu} \gamma^{\beta} 
+ g_{b \beta} \delta^{\mu}_{a} \gamma^{\nu} \gamma^{\beta}
] \tau^{c}
\psi \rangle. 
 \label{eq:VEV_comm_rot_tensor_third}
\end{align}
Thus, we obtain Table \ref{tab:relation_gamma-calc_rot_tensor}. 
Moreover, in the fourth equation of \eqref{eq:VEV_comm_rot_tensor}, since the isospin matrix $\tau^{c}$ does not appear in the commutation relations, it cannot be attributed to tensor condensation $\langle \bar{\psi} \gamma^{1} \gamma^{2} \tau^{3} \psi \rangle$. 
Thus, the rotational symmetry is not broken in the pseudotensor field channel. 
Therefore, if the tensor condensate realizes a nonzero value, namely $\langle i \bar{\psi} \gamma^{1} \gamma^{2} \tau^{3} \psi \rangle = \langle \bar{\psi} \Sigma_{3} \tau_{3} \psi \rangle \neq 0 $, the spatial rotational symmetry is spontaneously broken, and the following NG modes:
\begin{align}
	\begin{split}
\bar{\psi} \gamma^{2} \gamma^{3} \tau^{3} \psi &\approx t^{23}_{3}, 
\\ 
\bar{\psi} \gamma^{3} \gamma^{1} \tau^{3} \psi &\approx t^{31}_{3} 
	\end{split}
	\label{eq:NG_rotation}
\end{align}
appear.
Here, referring to the NG modes corresponding to the breaking of chiral symmetry \eqref{eq:NG_chiral}, we see that there are no modes identical to the NG modes corresponding to the breaking of spatial rotational symmetry \eqref{eq:NG_rotation}. 
Therefore, it is suggested that the mixing of symmetries, as shown in Sec. 
\ref{sec:level3-2}, does not occur in the breaking of chiral symmetry and space symmetry by tensor condensate.

\begin{table}[h]
 \begin{center}
   \caption{Vacuum expectation values (VEV) of commutation relations between spatial rotations and tensor fields are summarized.}
   \label{tab:relation_gamma-calc_rot_tensor}
  \begin{tabular}{ccccc}
  \hline \hline
  \multicolumn{2}{c}{} & \multicolumn{3}{c}{Channel} 
\\ \cline{3-5}
  \multicolumn{2}{c}{} & $\bar{\psi} \gamma^{1} \gamma^{2} \tau^{c} \psi$ & $\bar{\psi} \gamma^{2} \gamma^{3} \tau^{c} \psi$ & $\bar{\psi} \gamma^{3} \gamma^{1} \tau^{c} \psi$ 
\\ \hline
 Rot &  $i M_{12}$ & 0 & 0 & 0  
\\
 &  $i M_{23}$ & 0 & 0 & $ i \langle \bar{\psi} \gamma^{1} \gamma^{2} \tau^{c} \psi \rangle$ 
\\
  &  $i M_{31}$ & 0 & $ - i \langle \bar{\psi} \gamma^{1} \gamma^{2} \tau^{c} \psi \rangle $ & 0
\\
  \hline \hline
  \end{tabular}
 \end{center}
\end{table}

%%%%%%%%%%%%%%%%%%%%%%%%%%%%%%%%%%%%%
%%%%%%%%%%%%%%%%%%%%%%%%%%%%%%%%%%%%%

\subsection{\label{sec:level3-2}The case of the symmetry breaking by inhomogeneous chiral condensate}

In this subsection, let us consider the NG modes in the case where the condensate depends on the space coordinates.
The commutation relations between the generators of spatial translations $P_{a}$ and/or spatial rotations $M_{ab}$ and the fields $\bar{\psi} \psi$ and/or $\bar{\psi} i \gamma_{5} \tau^{c} \psi$ are obtained as follows:
\begin{align}
	\begin{split}
[iP_{a}, \bar{\psi} \psi]
&= \partial_{a} (\bar{\psi} \psi ), 
\\ 
[iP_{a}, \bar{\psi}  i \gamma_{5} \tau^{c} \psi]
&= \partial_{a} (\bar{\psi}  i \gamma_{5} \tau^{c}  \psi ), 
\\ 
[iM_{ab}, \bar{\psi} \psi]
&=
\partial_{b} ( x_{a} \bar{\psi} \psi ) 
- \partial_{a} ( x_{b} \bar{\psi} \psi ),  
\\ 
[iM_{ab}, \bar{\psi} i \gamma_{5} \tau^{c} \psi]
&=
\partial_{b} ( x_{a} \bar{\psi} i \gamma_{5} \tau^{c} \psi ) 
- \partial_{a} ( x_{b} \bar{\psi} i \gamma_{5} \tau^{c} \psi ). 
	\end{split}
\label{eq:comm_space_chiral}
\end{align}
We consider the inhomogeneous chiral condensate that depends only on the $z$-coordinate: 
\begin{align}
\langle \bar{\psi} \psi \rangle 
= \Delta \cos q z, 
\hspace{10pt}
\langle \bar{\psi} i \gamma_{5} \tau^{3} \psi \rangle 
= \Delta \sin q z. 
\label{eq:DCDW}
\end{align}
In this case, the vacuum expectation value (VEV) of the commutation relations with the generator of spatial symmetry \eqref{eq:comm_space_chiral} is given by
\begin{align}
	\begin{split}
\langle [iP_{a}, \bar{\psi} \psi] \rangle
&= \partial_{a} \langle \bar{\psi} \psi \rangle
= - \delta_{a3} q \Delta \sin q z, 
\\ 
\langle [iP_{a}, \bar{\psi} i \gamma_{5} \tau^{c} \psi] \rangle
&= \partial_{a} \langle \bar{\psi} i \gamma_{5} \tau^{c} \psi \rangle
= \delta_{a3} \delta^{c3} q \Delta \cos q z, 
\\ 
\langle [iM_{ab}, \bar{\psi} \psi] \rangle
&= \partial_{b} \Big( x_{a} \langle \bar{\psi} \psi \rangle \Big) - \partial_{a} \Big( x_{b} \langle \bar{\psi} \psi \rangle \Big) 
\\ 
&=
q \Delta 
\left[ \delta_{a3} x_{b} - \delta_{b3} x_{a} \right] \sin qz, 
\\ 
\langle [iM_{ab}, \bar{\psi} i \gamma_{5} \tau^{c} \psi] \rangle
&= \partial_{b} \Big( x_{a} \langle \bar{\psi} i \gamma_{5} \tau^{c} \psi \rangle \Big) - \partial_{a} \Big( x_{b} \langle \bar{\psi} i \gamma_{5} \tau^{c} \psi \rangle \Big) 
\\ 
&= - q \Delta \delta^{c3} \left[ \delta_{a3} x_{b} - \delta_{b3} x_{a} \right] \cos qz. 
	\end{split}
\label{eq:VEV_comm_space_chiral_inhomogeneous} 
\end{align} 
Moreover, from \eqref{eq:comm_relation_scalar}, the following equations hold for chiral transformations: 
\begin{align}
     \begin{split}
\langle [i Q_{A}^{a}, \bar{\psi} i \gamma_{5} \tau^{b} \psi] \rangle
&= - \delta^{ab} \langle \bar{\psi} \psi \rangle
= - \delta^{ab} \Delta \cos qz, 
\\ 
\langle  [i Q_{A}^{a}, \bar{\psi} \psi] \rangle 
&= \langle  \bar{\psi} i \gamma_{5} \tau^{a} \psi\rangle
= \delta^{a3} \Delta \sin qz, 
\\ 
\langle [i Q_{V}^{a}, \bar{\psi} i \gamma_{5} \tau^{b} \psi] \rangle 
&= \tensor{\epsilon}{^a^b^c} \langle \bar{\psi} i \gamma_{5} \tau^{c} \psi \rangle
= \tensor{\epsilon}{^a^b^3} \Delta \sin qz, 
\\ 
\langle [i Q_{V}^{a}, \bar{\psi} \psi] \rangle
&= 0
     \end{split}
\label{eq:VEV_comm_chiral_inhomogeneous}
\end{align}
From these, we obtain Table \ref{tab:VEV_comm_chiral_inhomogeneous}.

\begin{table}[h]
	\begin{center}
	\caption{VEV of commutation relations in inhomogeneous chiral condensate are summarized.}
	\label{tab:VEV_comm_chiral_inhomogeneous}
 	\begin{tabular}{cccccc} 
\hline \hline
    \multicolumn{2}{c}{} & \multicolumn{4}{c}{Channel} \\ \cline{3-6}
    \multicolumn{2}{c}{} &  $ \bar{\psi} \psi$  &  $\bar{\psi} i \gamma_{5} \tau^{1} \psi$  &  $\bar{\psi} i \gamma_{5} \tau^{2} \psi$  &  $\bar{\psi} i \gamma_{5} \tau^{3} \psi$ \\ \hline
Sym& $iP_{1}$ & 0 & 0 & 0 & 0  \\
   & $iP_{2}$ & 0 & 0 & 0 & 0  \\
   & $iP_{3}$ & $- q \Delta \sin qz$ & 0 & 0 & $q \Delta \cos qz$ \\ 
   & $iM_{12}$ & 0 & 0 & 0 & 0  \\
   & $iM_{23}$ & $-x_{2} q \Delta \sin qz$ & 0 & 0 &  $x_{2} q \Delta \cos qz$  \\
   & $iM_{31}$ & $x_{1} q \Delta \sin qz$ & 0 & 0 & $ - x_{1} q \Delta \cos qz$ \\ 
   & $iQ^{1}_{A}$ & 0 & $-\Delta \cos qz$ & 0 &  0  \\
   & $iQ^{2}_{A}$ & 0 & 0 & $-\Delta \cos qz$ &  0  \\
   & $iQ^{3}_{A}$ & $\Delta \sin qz$ & 0 & 0 &  $-\Delta \cos qz$  \\
   & $iQ^{1}_{V}$ & 0 & 0 & $\Delta \sin qz$ & 0 \\
   & $iQ^{2}_{V}$ & 0 & $- \Delta \sin qz$ & 0 &  0  \\
   & $iQ^{3}_{V}$ & 0 & 0 & 0 &  0  \\
\hline \hline
	\end{tabular}
	\end{center}
\end{table}
From Table \ref{tab:VEV_comm_chiral_inhomogeneous}, we can see the following relationship between the translation in the $z$ direction and the rotation in the plane containing the $z$-axis: 
\begin{align}
	\begin{split}
x_{2} \langle [iP_{3}, \Psi] \rangle
&=
\langle [iM_{23}, \Psi] \rangle
\\ 
- x_{1} \langle [iP_{3}, \Psi] \rangle
&=
\langle [iM_{31}, \Psi] \rangle
	\end{split}
	\label{eq:independent_massless_modes_space_symmetry}
\end{align} 
where $\Psi = \bar{\psi} \psi \ \mathrm{or}\  \bar{\psi} i \gamma_{5} \tau^{3} \psi$.
This is consistent with the assertion by Low and Manohar \cite{Low:2002sbs}, suggesting that $P_{a}$ and $M_{ab}$ do not give independent massless modes. 
On the other hand, it is evident that the two symmetries, translation $P_{3}$ and axial chiral transformation $Q^{3}_{A}$, are simultaneously broken due to the two condensates, $\langle \bar{\psi} \psi \rangle$ and $\langle \bar{\psi} i \gamma_{5} \tau^{3} \psi \rangle$. 
Rewriting these relations, we have
\begin{align}
	\begin{split}
\langle [iP_{3}, \bar{\psi} \psi]  \rangle &= \partial_{3} \langle \bar{\psi} \psi \rangle = - q \Delta \sin qz, 
\\ 
\langle  [iP_{3}, \bar{\psi} i \gamma_{5} \tau^{3} \psi]  \rangle &= \partial_{3} \langle  \bar{\psi} i \gamma_{5} \tau^{3} \psi\rangle = q \Delta \cos qz, 
\\ 
\langle [i Q_{A}^{3}, \bar{\psi} \psi] \rangle &= \langle  \bar{\psi} i \gamma_{5} \tau^{3} \psi\rangle = \Delta \sin qz, 
\\ 
\langle [i Q_{A}^{3}, \bar{\psi} i \gamma_{5} \tau^{3} \psi] \rangle &= - \delta^{33} \langle \bar{\psi} \psi \rangle = - \Delta \cos qz. 
	\end{split}
	\label{eq:VEV_2sym2cond}
\end{align}
As seen in Sec. \ref{sec:level4-3}, what may appear to be multiple NG modes are actually just one, as follows:
\begin{align}
\langle [i Q_{\mathrm{NG}}, \Phi_{\mathrm{NG}}] \rangle &= \sqrt{2}\Delta \ ,
\end{align}
where
\begin{align}
Q_{\mathrm{NG}} &= \frac{1}{\sqrt{{2}}} \left( Q_{A}^{3} - \frac{P_{3}}{{q}} \right) \ , \\
\Phi_{\mathrm{NG}} 
&= \sin qz \cdot (\bar{\psi} \psi) - \cos qz \cdot (\bar{\psi} i \gamma_{5} \tau^{3} \psi) .
\label{eq:PhiNG}
\end{align}
Details will be shown in Sec. \ref{sec:level4-3} with Appendix \ref{sec:appendixD}. 
A similar mixing of translational and internal symmetries is seen in Ref. \ \cite{Kobayashi:2014nrn}.

Considering all possible generator mixings and field mixings in the same way, we finally get the following three commutation relations for the independent NG modes:
\begin{align}
\begin{split}
\langle [iQ_{\mathrm{NG}}^{(1)}, \Phi_{\mathrm{NG}}^{(1)}] \rangle &= -\Delta \ ,\\
\langle [iQ_{\mathrm{NG}}^{(2)}, \Phi_{\mathrm{NG}}^{(2)}] \rangle &= -\Delta \ ,\\
\langle [iQ_{\mathrm{NG}}^{(3)}, \Phi_{\mathrm{NG}}^{(3)}] \rangle &= \Delta\sqrt{q^2(x_1^2+x_2^2)+2} \ ,
\end{split}
\end{align}
where
\begin{align}
\begin{split}
Q_{\mathrm{NG}}^{(1)} &= Q_A^1\cos qz + Q_V^2\sin qz \ , \\
Q_{\mathrm{NG}}^{(2)} &= Q_A^2\cos qz - Q_V^1\sin qz \ , \\
Q_{\mathrm{NG}}^{(3)} &= \frac{1}{\sqrt{q^2(x_1^2+x_2^2)+2}} \\
                                  &\qquad\times \left[ Q_A^3 -\frac{P_{3}}{q} -q(x_2M_{23} -x_1M_{31}) \right] \ ,
\end{split}
\end{align}
\begin{align}
\begin{split}
\Phi_{\mathrm{NG}}^{(1)} &= \bar{\psi} i \gamma_{5} \tau^{1} \psi \ ,\\
\Phi_{\mathrm{NG}}^{(2)} &= \bar{\psi} i \gamma_{5} \tau^{2} \psi \ ,\\
\Phi_{\mathrm{NG}}^{(3)} &= \sin qz \cdot (\bar{\psi} \psi) - \cos qz \cdot (\bar{\psi} i \gamma_{5} \tau^{3} \psi) \ .
\end{split}
\end{align}
Therefore, we obtain independent generators $Q_{\mathrm{NG}}^{(1)}$, $Q_{\mathrm{NG}}^{(2)}$, $Q_{\mathrm{NG}}^{(3)}$ and fields $\Phi_{\mathrm{NG}}^{(1)}$, $\Phi_{\mathrm{NG}}^{(2)}$, $\Phi_{\mathrm{NG}}^{(3)} $ corresponding to the three NG modes. This result reproduces the discussion of NG modes in Ref.\ \cite{Lee:2015lpi}.

%%%%%%%%%%%%%%%%%%%%%%%%%%%%%%%%%%%%%
%%%%%%%%%%%%%%%%%%%%%%%%%%%%%%%%%%%%%

\section{\label{sec:level4}Generalization}

Let us generalize the results obtained in the specific models in Sec. \ref{sec:level3}. 

\subsection{\label{sec:level4-1} The case of a single symmetry broken by two field operators}
We consider the case where a single generator $Q^{A}$ is broken by two field operators, $\Phi_{i}$ and $\Phi_{j}$: 
\begin{align}
	\begin{split}
\langle [i Q^{A}, \Phi_{i}] \rangle &= v_{p} \neq 0, 
\\ 
\langle [i Q^{A}, \Phi_{j}] \rangle &= v_{q} \neq 0. 
	\end{split}
	\label{eq:VEV_comm_gen_1by2}
\end{align}
Also, we assume that the fields $\Phi_{i}$ and $\Phi_{j}$ are orthogonal. 
Next, using $\Phi_{i}$ and $\Phi_{j}$, we redefine the fields $\Phi_{\mathrm{H}}$ and $\Phi_{\mathrm{NG}}$ as follows: 
\begin{align}
\begin{pmatrix}
\Phi_{\mathrm{H}} \\ \Phi_{\mathrm{NG}}
\end{pmatrix}
=
\begin{pmatrix}
\cos \theta & - \sin \theta \\ 
\sin \theta & \cos \theta
\end{pmatrix}
\begin{pmatrix}
\Phi_{i} \\ \Phi_{j}
\end{pmatrix}. 
\label{eq:Field_rotation_gen_1by2}
\end{align}
The fields $\Phi_{\mathrm{H}}$ and $\Phi_{\mathrm{NG}}$ constructed in this way remain orthogonal. 
In this case, if we take the rotation angle $\theta$ as
\begin{align}
\sin \theta = \frac{v_{p}}{\sqrt{v_{p}^{2} + v_{q}^{2}}}, \hspace{8mm}
\cos \theta = \frac{v_{q}}{\sqrt{v_{p}^{2} + v_{q}^{2}}},
\label{eq:angle_gen_1by2}
\end{align}
then $\Phi_{\mathrm{H}}$ and $\Phi_{\mathrm{NG}}$ are expressed as follows: 
\begin{align}
\Phi_{\mathrm{H}} 
= \frac{v_{q}}{\sqrt{v_{p}^{2} + v_{q}^{2}}} \Phi_{i} 
- \frac{v_{p}}{\sqrt{v_{p}^{2} + v_{q}^{2}}} \Phi_{j}, 
\label{eq:Higgs_mode_gen_1by2}
\\ 
\Phi_{\mathrm{NG}} 
= \frac{v_{p}}{\sqrt{v_{p}^{2} + v_{q}^{2}}} \Phi_{i} 
+ \frac{v_{q}}{\sqrt{v_{p}^{2} + v_{q}^{2}}} \Phi_{j}. 
\label{eq:NG_mode_gen_1by2}
\end{align}
At this point, since they satisfy
\begin{align}
	\begin{split}
\langle [iQ^{A}, \Phi_{\mathrm{H}}] \rangle 
&= 0, 
\\ 
\langle [iQ^{A}, \Phi_{\mathrm{NG}}] \rangle
&= \sqrt{ v_{p}^{2} + v_{q}^{2} }
 \neq 0,
	\end{split}
	\label{eq:SSB_gen_1by2}
\end{align}
even though it might initially appear that two NG modes emerge from \eqref{eq:VEV_comm_gen_1by2}, in reality, $\Phi_{\mathrm{NG}}$ appears as a single NG mode.

Next, we revisit the discussion from Sec. \ref{sec:level3-1} in light of this generalization. 
From \eqref{eq:SSB_QA3}, the following equations hold:
\begin{align}
     \begin{split}
\langle [i Q_{A}^{a = 3}, \pi^{3}]  \rangle 
&=  
- \sigma_{0} \neq 0, 
\\ 
\langle [i Q_{A}^{a = 3}, \varpi^{12}_{h}]  \rangle
&= - t_{0} \neq 0. 
     \end{split}
\label{eq:SSB_QA3_1by2}
\end{align}
This corresponds to \eqref{eq:VEV_comm_gen_1by2} with $Q^{A} = Q^{a = 3}_{A}$, $\Phi_{i} = \pi^{3}, \Phi_{j} = \varpi^{12}_{h}$, $v_{p} = - \sigma_{0}$ and $v_{q} = - t_{0}$. 
Substituting these into \eqref{eq:Higgs_mode_gen_1by2} and \eqref{eq:NG_mode_gen_1by2}, we can construct $\Phi_{\mathrm{H}}$ and $\Phi_{\mathrm{NG}}$ as follows: 
\begin{align}
\Phi_{\mathrm{H}} 
= \frac{ - t_{0} \pi^{3} - \sigma_{0} \varpi^{12}_{h} }{\sqrt{\sigma_{0}^{2} + t_{0}^{2}}}, 
\label{eq:Higgs_mode_eg_1by2}
\\ 
\Phi_{\mathrm{NG}} 
= -\frac{\sigma_{0} \pi^{3} + t_{0} \varpi^{12}_{h} }{\sqrt{\sigma_{0}^{2} + t_{0}^{2}}}. 
\label{eq:NG_mode_eg_1by2}
\end{align}
In this case, $\Phi_{\mathrm{H}}$ and $\Phi_{\mathrm{NG}}$ satisfy \eqref{eq:SSB_gen_1by2}. 
This result is consistent with the discussion in Sec. \ref{sec:level3-1}. 
Therefore, Sec. \ref{sec:level3-1} can be seen as an example of the aforementioned generalization.

%%%%%%%%%%%%%%%%%%%%%%%%%%%%%%%%%%%%%
%%%%%%%%%%%%%%%%%%%%%%%%%%%%%%%%%%%%%

\subsection{\label{sec:level4-2} The case of two symmetries broken by a single field operator}

In this section, we consider the case where two generators, $Q^{A}$ and $Q^{B}$, are broken by a single field operator $\Phi$:
\begin{align}
	\begin{split}
\langle [i Q^{A}, \Phi] \rangle &= v_{p} \neq 0, 
\\ 
\langle [i Q^{B}, \Phi] \rangle &= v_{q} \neq 0. 
	\end{split}
\label{eq:SSB_2sym1field}
\end{align}
Now, the generators $Q^{A}$ and $Q^{B}$ are mixed and redefined as $Q_{\mathrm{H}}$ and $Q_{\mathrm{NG}}$ as follows:
\begin{align}
\begin{pmatrix}
Q_{\mathrm{H}} \\ Q_{\mathrm{NG}}
\end{pmatrix}
=
\begin{pmatrix}
\cos \theta & - \sin \theta \\ 
\sin \theta & \cos \theta
\end{pmatrix}
\begin{pmatrix}
Q^{A} \\ Q^{B}
\end{pmatrix}.
\label{eq:redefine_2sym1field}
\end{align}
Here, let us take the rotation angle $\theta$ as 
\begin{align}
\cos \theta = \frac{v_{q}}{\sqrt{ v_{p}^{2} + v_{q}^{2} }}, \hspace{8mm}
\sin \theta = \frac{v_{p}}{\sqrt{ v_{p}^{2} + v_{q}^{2} }}. 
\label{eq:angle_2sym1field}
\end{align}
By substituting these into \eqref{eq:redefine_2sym1field}, we obtain 
\begin{align}
Q_{\mathrm{H}} &= \frac{v_{q}}{\sqrt{ v_{p}^{2} + v_{q}^{2} }} \cdot Q^{A} - \frac{v_{p}}{\sqrt{ v_{p}^{2} + v_{q}^{2} }} \cdot Q^{B}, 
\label{eq:unbroken_sym_2sym1field}
\\ 
Q_{\mathrm{NG}} &= \frac{v_{p}}{\sqrt{ v_{p}^{2} + v_{q}^{2} }} \cdot Q^{A} + \frac{v_{q}}{\sqrt{ v_{p}^{2} + v_{q}^{2} }} \cdot Q^{B}. 
\label{eq:broken_sym_2sym1field}
\end{align}
In this case, Eq. \eqref{eq:SSB_2sym1field} can be rewritten as
\begin{align}
	\begin{split}
\langle [i Q_{\mathrm{H}}, \Phi] \rangle &= 0, 
\\ 
\langle [i Q_{\mathrm{NG}}, \Phi] \rangle &= \sqrt{ v_{p}^{2} + v_{q}^{2} } \neq 0. 
	\end{split}
\label{eq:SSB_resym_2sym1field}
\end{align}
Even though it might initially appear that two broken generators emerge from \eqref{eq:SSB_2sym1field}, in reality, $Q_{\mathrm{NG}}$ appears as a single broken generator.

%%%%%%%%%%%%%%%%%%%%%%%%%%%%%%%%%%%%%
%%%%%%%%%%%%%%%%%%%%%%%%%%%%%%%%%%%%%

\subsection{\label{sec:level4-3} The case of two symmetries broken by two field operators}

Next, we consider the case where two generators, $Q^{A}$ and $Q^{B}$, are broken by two field operators, $\Phi_{i}$ and $\Phi_{j}$.  
\begin{align}
	\begin{split}
\langle [i Q^{A}, \Phi_{i}] \rangle &= v_{p} \neq 0, 
\\ 
\langle [i Q^{A}, \Phi_{j}] \rangle &= v_{q} \neq 0, 
\\ 
\langle [i Q^{B}, \Phi_{i}] \rangle &= \alpha v_{p} \neq 0, 
\\ 
\langle [i Q^{B}, \Phi_{j}] \rangle &= \beta v_{q} \neq 0. 
	\end{split}
\label{eq:SSB_2sym2cond}
\end{align}
In the following, we mix the broken generators and the field operators in the same way as for Secs. \ref{sec:level4-2} and \ref{sec:level4-1}, respectively. 
However, for $\alpha \neq \beta$, the mixing is not uniquely determined because there are two ways to mix the generators: one focusing on $\Phi_{i}$ and based on the first and third equations of \eqref{eq:SSB_2sym2cond}, and the other focusing on $\Phi_{j}$ and based on the second and fourth equations of \eqref{eq:SSB_2sym2cond}. 
Therefore, we restrict our discussion to the case of $\alpha = \beta$.

We mix the generators $Q^{A}$ and $Q^{B}$, and define them as $Q_{\mathrm{H}}$ and $Q_{\mathrm{NG}}$ as follows:
\begin{align}
Q_{\mathrm{H}} &= \frac{\alpha}{\sqrt{1 + \alpha^{2}}} \cdot Q^{A} - \frac{1}{\sqrt{1 + \alpha^{2}}} \cdot Q^{B}, 
\label{eq:unbroken_sym_2sym2cond}
\\ 
Q_{\mathrm{NG}} &= \frac{1}{\sqrt{1 + \alpha^{2}}} \cdot Q^{A} + \frac{\alpha}{\sqrt{1 + \alpha^{2}}} \cdot Q^{B}. 
\label{eq:broken_sym_2sym2cond}
\end{align}
In this case, Eq. \eqref{eq:SSB_2sym2cond} can be rewritten as
\begin{align}
	\begin{split}
\langle [i Q_{\mathrm{H}}, \Phi_{i}] \rangle &= 0, 
\\ 
\langle [i Q_{\mathrm{H}}, \Phi_{j}] \rangle &= 0, 
\\ 
\langle [i Q_{\mathrm{NG}}, \Phi_{i}] \rangle &= \sqrt{1 + \alpha^{2}} v_{p} \neq 0, 
\\ 
\langle [i Q_{\mathrm{NG}}, \Phi_{j}] \rangle &= \sqrt{1 + \alpha^{2}} v_{q}\neq 0. 
	\end{split}
\label{eq:SSB_resym_angle_2sym2cond}
\end{align}
That is, we can divide into the broken generator $Q_{\mathrm{NG}}$ and the unbroken generator $Q_{\mathrm{H}}$. 
Here, we mix the field operators, $\Phi_{i}$ and $\Phi_{j}$, and define $\Phi_{\mathrm{H}}$ and $\Phi_{\mathrm{NG}}$ as follows: 
\begin{align}
\Phi_{\mathrm{H}} 
&= \frac{v_{q}}{\sqrt{v_p^2 + v_q^2}} \Phi_{i} - \frac{v_{p}}{\sqrt{v_p^2 + v_q^2}} \Phi_{j}, 
\label{eq:Higgs_mode_2sym2cond}
\\ 
\Phi_{\mathrm{NG}} 
&= \frac{v_{p}}{\sqrt{v_p^2 + v_q^2}} \Phi_{i} + \frac{v_{q}}{\sqrt{v_p^2 + v_q^2}} \Phi_{j}.
\label{eq:NG_mode_2sym2cond}
\end{align}
In this case, from \eqref{eq:SSB_resym_angle_2sym2cond}, we have
\begin{align}
	\begin{split}
\langle [i Q_{\mathrm{H}}, \Phi_{\mathrm{H}}] \rangle &= 0, 
\\ 
\langle [i Q_{\mathrm{H}}, \Phi_{\mathrm{NG}}] \rangle &= 0, 
\\ 
\langle [iQ_{\mathrm{NG}}, \Phi_{\mathrm{H}}] \rangle 
&= \sqrt{\frac{1+\alpha^2}{v_p^2+v_q^2}}(v_{p} v_{q} - v_{p}v_{q} ) = 0, 
\\ 
\langle [iQ_{\mathrm{NG}}, \Phi_{\mathrm{NG}}] \rangle
&= \sqrt{1+\alpha^2}\sqrt{v_{p}^{2} + v_{q}^{2}} \neq 0. 
	\end{split}
\label{eq:SSB_final_2sym2cond}
\end{align}
Thus, $Q_{\mathrm{H}}$ is unbroken, and for the broken $Q_{\mathrm{NG}}$, $\Phi_{\mathrm{H}}$ is the Higgs mode, and $\Phi_{\mathrm{NG}}$ is the NG mode.

Finally, we revisit the discussion from Sec. \ref{sec:level3-2} in light of this generalization. 
From \eqref{eq:VEV_2sym2cond}, the following equations hold:
\begin{align}
	\begin{split}
\langle [i Q_{A}^{3}, \bar{\psi} \psi] \rangle &= \Delta \sin qz, 
\\ 
\langle [i Q_{A}^{3}, \bar{\psi} i \gamma_{5} \tau^{3} \psi] \rangle &= - \Delta \cos qz, 
\\ 
\langle [isP_{3}, \bar{\psi} \psi]  \rangle &= - sq \Delta \sin qz, 
\\ 
\langle  [isP_{3}, \bar{\psi} i \gamma_{5} \tau^{3} \psi]  \rangle &= sq \Delta \cos qz, 
	\end{split}
\label{eq:SSB_eg_2sym2cond}
\end{align}
where, to align the dimensions of $P_{3}$ with $Q_{A}^{3}$, we introduce a parameter $s$ that has the dimension of length. 
The only dimensional parameter except for the parameters appearing in the model Lagrangian initially is $q$, which is introduced in the condensate as a wave number of a modulation.
Therefore, in this case, the parameter $s$ with the dimension of length should be proportional to the dimensional parameter $1/q$. 
Thus, let us adopt $-1/q$ as $s$ because the chiral rotation $Q_A^3$ and the translation $-P_3/q$ give the same transformation for the inhomogeneous condensate as is discussed in Appendix \ref{sec:appendixD}. 
Here, let us regard the general quantities $Q^A, Q^B, \Phi_i, \Phi_j, v_p$ and $v_q$ as follows:
\begin{gather}
	\begin{split}
Q^{A} = Q_{A}^{3},& \hspace{8pt} Q^{B} = 
-\frac{P_{3}}{q}, 
\\ 
\Phi_{i} = \bar{\psi} \psi,& \hspace{8pt} \Phi_{j} = \bar{\psi} i \gamma_{5} \tau^{3} \psi , 
\\ 
v_{p} = \Delta \sin qz,& \hspace{8pt} v_{q} = - \Delta \cos qz. 
	\end{split}
\label{eq:replace_eg_2sym2cond}
\end{gather}
Then, Eq.\eqref{eq:SSB_eg_2sym2cond} can be rewritten as follows:
\begin{align}
	\begin{split}
\langle [i Q^{A}, \Phi_{i}] \rangle &= v_{p}, 
\\ 
\langle [i Q^{A}, \Phi_{j}] \rangle &= v_{q}, 
\\ 
\langle [iQ^{B}, \Phi_{i}]  \rangle &=   {v_{p}}, 
\\ 
\langle  [iQ^{B}, \Phi_{j}]  \rangle &=  {v_{q}}.
	\end{split}
\end{align}
These correspond to the equations in \eqref{eq:SSB_2sym2cond} with $\alpha = \beta = -sq=1$. 
Therefore, the general theory for the case of $\alpha = \beta$ can be applied. 
By substituting \eqref{eq:replace_eg_2sym2cond} into \eqref{eq:unbroken_sym_2sym2cond} and \eqref{eq:broken_sym_2sym2cond}, we obtain the unbroken generator $Q_{\mathrm{H}}$ and the broken generator $Q_{\mathrm{NG}}$ as follows:
\begin{align}
Q_{\mathrm{H}} &=  \frac{{1}}{\sqrt{{2}}} \left( Q_{A}^{3} +  \frac{P_{3}}{{q}} \right), 
\label{eq:unbroken_sym_eg_2sym2cond}
\\ 
Q_{\mathrm{NG}} &= \frac{1}{\sqrt{{2}}} \left( Q_{A}^{3} - \frac{P_{3}}{{q}} \right). 
\label{eq:broken_sym_eg_2sym2cond}
\end{align}
Similarly, by substituting \eqref{eq:replace_eg_2sym2cond} into \eqref{eq:Higgs_mode_2sym2cond} and \eqref{eq:NG_mode_2sym2cond}, we obtain the Higgs mode $\Phi_{\mathrm{H}}$  and the NG mode $\Phi_{\mathrm{NG}}$  as follows:
\begin{align}
\Phi_{\mathrm{H}} 
&= - \left( \cos qz \cdot (\bar{\psi} \psi) + \sin qz \cdot (\bar{\psi} i \gamma_{5} \tau^{3} \psi) \right), 
\label{eq:Higgs_mode_eg_2sym2cond}
\\ 
\Phi_{\mathrm{NG}} 
&= \sin qz \cdot (\bar{\psi} \psi) - \cos qz \cdot (\bar{\psi} i \gamma_{5} \tau^{3} \psi) . 
\label{eq:NG_mode_eg_2sym2cond}
\end{align}
In this case, $\Phi_{\mathrm{H}}$ and $\Phi_{\mathrm{NG}}$ satisfy \eqref{eq:SSB_final_2sym2cond}. 
Therefore, in \eqref{eq:VEV_2sym2cond}, it is found that $Q_{\mathrm{NG}}$ is broken, resulting in the emergence of a single NG mode, $\Phi_{\mathrm{NG}}$. 
Here, $\Phi_{\rm NG}$ is nothing else (\ref{eq:PhiNG}).
%%%%%%%%%%%%%%%%%%%%%%%%%%%%%%%%%%%%%
%%%%%%%%%%%%%%%%%%%%%%%%%%%%%%%%%%%%%

\subsection{\label{sec:level4-4} The rules of a single symmetry breaking by multiple field operators}

In the following, we describe the rules for the case where $N$ field operators break a single symmetry. 

Assume that a single symmetry is broken by $N$ field operators as follows:
\begin{align}
	\begin{split}
\langle [i Q^{A}, \Phi_{1}] \rangle&= v_{1} \neq 0, \\ 
\langle [i Q^{A}, \Phi_{2}] \rangle &= v_{2} \neq 0, \\ 
&\vdots \\ 
\langle [i Q^{A}, \Phi_{N}] \rangle &= v_{n} \neq 0. 
	\end{split}
	\label{eq:SSB_gen_1symNcond}
\end{align}
From the discussions in Sec. \ref{sec:level4-1} and Appendix \ref{sec:appendixC} on the cases where $N = 3, 4$ condensates break a single symmetry, the following three rules can be inferred: 
The first rule is that by appropriately rotating the $N$ orthogonal modes, one NG mode and $N-1$ Higgs modes can be obtained. 
The second rule is that the NG mode is determined by $N-1$ rotation parameters. 
This is because the rotation of $N-1$ modes results in ${}_N C_{2}$ parameters, of which ${}_{N-1} C_{2}$ are arbitrary rotations related to the Higgs modes, leading to ${}_{N}C_{2} - {}_{N-1}C_{2} = N -1$ rotation parameters that determine the NG mode. 
Here, $_{N} C_{r}$ is the binomial coefficient, representing the number of combinations:
\begin{align}
_{N} C_{r} = 
\begin{pmatrix} 
N \\ r 
\end{pmatrix}
= 
\frac{N!}{r!(N-r)!}. 
\label{eq:binomial coeeficient}
\end{align}
In this context, we use the case with $r = 2$. 
The third rule is that the condensate leading to the NG mode is denoted by $v = \sqrt{ \sum_{i=1}^{N} v_{i}^{2} }$. 
This relationship determines the unique parameters $(v, \theta_{1}, \cdots, \theta_{N-1})$ and the original mode condensates $(v_{1}, v_{2}, \cdots, v_{N})$ that satisfy this equation.
By rotating the $N$ modes according to these rules, one NG mode and $N-1$ Higgs modes can be obtained as follows:
\begin{equation}
\Phi_{\mathrm{NG}} = \frac{v_{1}}{v} \Phi_{1} + \frac{v_{2}}{v} \Phi_{2} + \cdots + \frac{v_{N}}{v} \Phi_{N}, 
	\label{eq:NG_gen_1symNcond}
\end{equation}
\begin{align}
	\begin{split}
\langle [i Q^{A}, \Phi_{\mathrm{NG}}] \rangle &= v \neq 0, \\ 
\langle [i Q^{A}, \Phi_{\mathrm{H}_{1}}] \rangle &= 0, \\ 
&\vdots \\ 
\langle [i Q^{A}, \Phi_{\mathrm{H}_{N-1}}] \rangle &= 0. 
	\end{split}
	\label{eq:Higgs_gen_1symNcond}
\end{align}
Note that $\Phi_{\mathrm{H}_{1}}, \cdots, \Phi_{\mathrm{H}_{N-1}}$ are determined by ${}_{N-1} C_{2}$ arbitrary parameters, and therefore, it is not uniquely defined.

Finally, we mention the rules for mixing generators when breaking multiple generators $Q_{1}, \cdots, Q_{N}$ in the common field operator $\Phi$. 
That is, 
\begin{align}
	\begin{split}
\langle [iQ_{1}, \Phi] \rangle &= v_{1} \neq 0, 
\\ 
\langle [iQ_{2}, \Phi] \rangle &= v_{2} \neq 0, 
\\ 
&\vdots
\\ 
\langle [iQ_{N}, \Phi] \rangle &= v_{N} \neq 0. 
	\end{split}
	\label{eq:SSB_gen_Nsym1field}
\end{align}
Here, $v_{1}, \ldots, v_{N}$ do not necessarily need to be different condensates. As seen in Sec. \ref{sec:level4-2}, generators can be mixed similarly to the mixing of modes. 
Therefore, the aforementioned rules can also be applied to the mixing of $Q_{1}, \ldots, Q_{N}$. 
Specifically, the following three rules are inferred. 
The first rule is that by appropriately rotating the $N$ generators, one broken generator $Q_{\mathrm{NG}}$ and $N-1$ unbroken generators $Q_{\mathrm{H}_{i}}$ with $i = 1, 2, \ldots, N-1$ can be obtained.
The second rule is that the broken generator $Q_{\mathrm{NG}}$ is determined by $N-1$ rotation parameters. 
Moreover, $N-1$ unbroken generators $Q_{\mathrm{H}_{i}}$ results in ${}_N C_{2}$ parameters, of which ${}_{N-1} C_{2}$ are arbitrary rotations. 
The third rule is that the condensate leading to the broken generator $Q_{\mathrm{NG}}$ is denoted by $v = \sqrt{ \sum_{i=1}^{N} v_{i}^{2} }$. 
This relationship determines the unique parameters $(v, \theta_{1}, \cdots, \theta_{N-1})$ and the original mode condensates $(v_{1}, v_{2}, \cdots, v_{N})$ that satisfy the next equation: 
\begin{equation}
i Q_{\mathrm{NG}} = \frac{v_{1}}{v} i Q_{1} + \frac{v_{2}}{v} i Q_{2} + \cdots + \frac{v_{N}}{v} i Q_{N}. 
	\label{eq:NG_gen_Nsym1field}
\end{equation}
%%%%%%%%%%%%%%%%%%%%%%%%%%%%%%%%%%%%%
%%%%%%%%%%%%%%%%%%%%%%%%%%%%%%%%%%%%%
\section{\label{sec:level5}summary}

In this paper, we present general rules for determining the broken generators and NG modes in the case of symmetry breaking by multiple condensates through knowledge of the symmetry breaking pattern on the exotic chiral condensates. 
Then, at first glance, the number of broken generators does not match the number of NG modes. 
It is necessary that the linear combinations of broken generators and those of elementary fields should be considered to establish a one-to-one correspondence.
These rules are based on the method using algebraic commutation relations. 

We considered specific examples of tensor-type condensates arising from tensor interaction, and the inhomogeneous chiral condensates as exotic chiral condensates. 
First, for the tensor-type condensate, we discussed two methods to derive the NG modes. 
The first method involves a use of the two-point vertex function. 
By employing this method, we observed that the presence of a tensor-type condensate breaks the chiral symmetry, resulting in two types of NG modes associated with the breaking of $su_{A}(2)$ chiral symmetry. 
By diagonalizing the meson mass matrix, we found that meson mixing occurs which leads to the NG mode and the Higgs mode. 
The second method was given by the use of algebraic commutation relations. 
Based on this method, we also considered the breaking of spatial symmetry in addition to the chiral symmetry breaking. 
This allowed us to completely reproduce the results obtained by the method using the two-point vertex function for the breaking of chiral symmetry and showed that the breaking of spatial symmetry and chiral symmetry are independent.

Second, for the dual chiral density wave as an inhomogeneous chiral condensate, we investigated the pattern of symmetry breaking based on the method of algebraic commutation relations and then discussed the results based on the derived general rules. 
The discussion using algebraic commutation relations revealed that both the chiral symmetry and the spatial symmetry are broken, and the NG modes associated with the breakings of $su_{A}(2)$ chiral symmetry, spatial translational symmetry, and spatial rotational symmetry, which lead to the same transformations for the dual chiral density wave, are nonindependent. 
Finally, we obtain the commutation relations corresponding to three independent NG modes, whose constituent generators and fields can be expressed as linear combinations of the original generators and fields.

Investigating the mathematical background of the derived general rules and conditions under which internal symmetry and spatial symmetry can be identified as the same is one of the main goals of our future work.

%%%%%%%%%%%%%%%%%%%%%%%%%%%%%%%%%%%%%
\begin{acknowledgments}

The authors acknowledge the members of the Nuclear Theory Group in Kochi University, especially to Professor K.\ Iida and Professor E.\ Nakano for useful comments.

\end{acknowledgments}
%%%%%%%%%%%%%%%%%%%%%%%%%%%%%%%%%%%%%

%%%%%%%%%%%%%%%%%%%%%%%%%%%%%%%%%%%%%
%%%%%%%%%%%%%%%%%%%%%%%%%%%%%%%%%%%%%
\appendix

\section{\label{sec:appendixA}Derivation of the gap equation}

Noting that $\left\{\gamma^{\mu}, \gamma^{\nu} \right\} = 2 g^{\mu \nu}$ and $(AB)^{-1} = B^{-1} A^{-1}$, we transform the trace part of \eqref{eq:gap_chiral1} as follows:
\begin{align}
&\ \tr \left[ 
\frac{1}{\cancel{k} - 2 \sigma_{0} - 4 i t_{0} \gamma^{1} \gamma^{2} \tau_{3} } 
\right]
\notag \\ 
&=
\tr \Bigg[ 
\frac{1}{\displaystyle \left\{ \cancel{k} + \left( 2 \sigma_{0} + 4 i t_{0} \gamma^{1} \gamma^{2} \tau_{3}\right) \right\}  \left\{ \cancel{k} - \left( 2 \sigma_{0} + 4 i t_{0} \gamma^{1} \gamma^{2} \tau_{3}\right) \right\} } 
\notag \\
&\hspace{8pt} \times 
\left\{ \cancel{k} + \left( 2 \sigma_{0} + 4 i t_{0} \gamma^{1} \gamma^{2} \tau_{3} \right) \right\}
\Bigg]
\notag \\ 
&=  
\tr \, \Biggl[ 
\frac{1}{\displaystyle \left( k^{2} - 4 \sigma_{0}^{2} - 16 t_{0}^{2} \right) - 8 i t_{0} \tau_{3} \left( k_{1} \gamma^{2} - k_{2} \gamma^{1} - 2 \sigma_{0} \gamma^{1} \gamma^{2} \right) }  
\notag \\ 
&\hspace{4pt} \times 
\frac{1}{ \displaystyle \left( k^{2} - 4 \sigma_{0}^{2} - 16 t_{0}^{2} \right) + 8 i t_{0} \tau_{3} \left( k_{1} \gamma^{2} - k_{2} \gamma^{1} - 2 \sigma_{0} \gamma^{1} \gamma^{2} \right) }
\notag \\ 
&\hspace{4pt} \times 
\left\{ \left( k^{2} - 4 \sigma_{0}^{2} + 16 t_{0}^{2} \right) - 8 i t_{0} \tau_{3} \left( k_{1} \gamma^{2} - k_{2} \gamma^{1} - 2 \sigma_{0} \gamma^{1} \gamma^{2} \right) \right\} 
\notag \\ 
& \hspace{4pt} \times 
\left\{ \cancel{k} + \left( 2 \sigma_{0} + 4 i t_{0} \gamma^{1} \gamma^{2} \tau_{3} \right) \right\}
\Bigg]
\notag \\ 
&= 
\frac{1}{ \left( k^{2} - 4 \sigma_{0}^{2} - 16 t_{0}^{2} \right)^{2} - 64 t_{0}^{2} \left( k_{1}^{2} + k_{2}^{2} + 4 \sigma_{0}^{2} \right) }
\notag \\ 
&\hspace{8pt} \times 
\tr \left[
\left\{ \left( k^{2} - 4 \sigma_{0}^{2} - 16 t_{0}^{2} \right) \right. \right. \notag \\
&\hspace{40pt} \left.- 8 i t_{0} \tau_{3} \left( k_{1} \gamma^{2} - k_{2} \gamma^{1} - 2 \sigma_{0} \gamma^{1} \gamma^{2} \right) \right\} 
\notag \\ 
&\hspace{28pt} \times \left.
\left\{ \cancel{k} + \left( 2 \sigma_{0} + 4 i t_{0} \gamma^{1} \gamma^{2} \tau_{3} \right) \right\}
\right]. 
\label{eq:gap_chiral2_transform1}
\end{align}
Here, note that the trace is taken over the Dirac gamma matrices, isospin space, and color space. Next, using the trace formula
\begin{align}
\begin{split}
 \tr \, \tau_{i} &= 0, 
\\ 
 \tr \, \tau_{i}^{2} &= 2, 
\\ 
 \tr \, (\gamma^{\mu} \gamma^{\nu}) &= 4 \tensor{g}{^\mu^\nu}, 
\\ 
  \tr \, (\gamma^{\mu_{1}} \cdots \gamma^{\mu_{n}}) &= 0 \hspace{4pt} (n : \text{odd number})
\end{split}
\label{eq:trace_formula}
\end{align}
we calculate the trace part of \eqref{eq:gap_chiral2_transform1} to obtain
\begin{align}
& \tr \left[
\left\{ \left( k^{2} - 4 \sigma_{0}^{2} - 16 t_{0}^{2} \right) - 8 i t_{0} \tau_{3} \left( k_{1} \gamma^{2} - k_{2} \gamma^{1} - 2 \sigma_{0} \gamma^{1} \gamma^{2} \right) \right\} \right. 
\notag \\ 
& \hspace{12pt} \times \left.
\left\{ \cancel{k} + \left( 2 \sigma_{0} + 4 i t_{0} \gamma^{1} \gamma^{2} \tau_{3} \right) \right\}
\right]
\notag \\ 
&= 48 \sigma_{0} \left\{ \left( k^{2} - 4 \sigma_{0}^{2} - 16 t_{0}^{2} \right) + 32 t_{0}^{2} \right\}. 
\label{eq:gap_chiral2_transform2}
\end{align}
Therefore, Eq.\eqref{eq:gap_chiral1} becomes
\begin{align}
&\sigma_{0} \left(  1 + 48 G_{S} \int \frac{\dd^{4} k}{i (2\pi)^{4}}  \right.
\notag \\ 
&\hspace{12pt} \times \left.
\frac{ k^{2} - 4 \sigma_{0}^{2} + 16 t_{0}^{2} }{ \displaystyle (k^{2} - 4 \sigma_{0}^{2} - 16 t_{0}^{2})^{2} - 64 t_{0}^{2} \left( k_{1}^{2} + k_{2}^{2} + 4 \sigma_{0}^{2} \right) }  \right)
= 0.
\label{eq:gap_chiral2_transform3}
\end{align}
This equation is identical to Eq. \eqref{eq:gap_chiral2}. 

Similarly, by transforming \eqref{eq:gap_tensor1}, we obtain \eqref{eq:gap_tensor2}.

%%%%%%%%%%%%%%%%%%%%%%%%%%%%%%%%%%%%%
%%%%%%%%%%%%%%%%%%%%%%%%%%%%%%%%%%%%%

\section{\label{sec:appendixB}Calculation of the two-point vertex functions}

Transform the trace part of \eqref{eq:tpvf_p0_pi3} in the same way as in the Appendix \ref{sec:appendixA}:
\begin{align}
&\ \tr \Bigg[
(-2i\gamma_{5} \tau_{3}) \frac{1}{\cancel{k} - 2 \sigma_{0} - 4 i t_{0} \gamma^{1} \gamma^{2} \tau_{3} }
\notag \\ 
& \hspace{24pt} \times 
(-2i\gamma_{5} \tau_{3}) \frac{1}{\cancel{k} - 2 \sigma_{0} - 4 i t_{0} \gamma^{1} \gamma^{2} \tau_{3} }
\Bigg]
\notag \\ 
&=
\tr \left[ -
\gamma_{5} \frac{1}{\cancel{k} - 2 \sigma_{0} - 4 i t_{0} \gamma^{1} \gamma^{2} \tau_{3} }
\gamma_{5} \frac{1}{\cancel{k} - 2 \sigma_{0} - 4 i t_{0} \gamma^{1} \gamma^{2} \tau_{3} }
\right]
\notag \\ 
&=
\tr \left[ 
\frac{-1}{ \{ \cancel{k} + ( 2 \sigma_{0} + 4 i t_{0} \gamma^{1} \gamma^{2} \tau_{3} ) \} \{\cancel{k} - ( 2 \sigma_{0} + 4 i t_{0} \gamma^{1} \gamma^{2} \tau_{3} ) \} }
\right]
\notag \\ 
&=
\tr \left[
\frac{-1}{ (k^{2} - 4 \sigma_{0}^{2} - 16 t_{0}^{2}) + 8 i t_{0} \tau_{3} (k_{1} \gamma^{2} - k_{2} \gamma^{1} - 2 \sigma_{0} \gamma^{1} \gamma^{2}) }
\right]
\notag \\ 
&=
\tr \left[
\frac{-1}{ (k^{2} - 4 \sigma_{0}^{2} - 16 t_{0}^{2}) - 8 i t_{0} \tau_{3} (k_{1} \gamma^{2} - k_{2} \gamma^{1} - 2 \sigma_{0} \gamma^{1} \gamma^{2}) } 
\right.
\notag \\ 
& \times 
\frac{1}{ (k^{2} - 4 \sigma_{0}^{2} - 16 t_{0}^{2}) + 8 i t_{0} \tau_{3} (k_{1} \gamma^{2} - k_{2} \gamma^{1} - 2 \sigma_{0} \gamma^{1} \gamma^{2}) } 
\nonumber \\ 
& \times \left. \{ (k^{2} - 4 \sigma_{0}^{2} - 16 t_{0}^{2}) - 8 i t_{0} \tau_{3} (k_{1} \gamma^{2} - k_{2} \gamma^{1} - 2 \sigma_{0} \gamma^{1} \gamma^{2}) \}
\right]
\notag \\ 
&= -
\frac{1}{ (k^{2} - 4 \sigma_{0}^{2} - 16 t_{0}^{2})^{2} - 64t_{0}^{2} (k_{1}^{2} + k_{2}^{2} + 4 \sigma_{0}^{2}) }
\notag \\
& \hspace{1pt} \times 
\tr 
\left[
(k^{2} - 4 \sigma_{0}^{2} - 16 t_{0}^{2}) - 8 i t_{0} \tau_{3} (k_{1} \gamma^{2} - k_{2} \gamma^{1} - 2 \sigma_{0} \gamma^{1} \gamma^{2})
\right].
\label{eq:tpvf_pi3-pi3_transform1}
\end{align}
Deriving the trace part of \eqref{eq:tpvf_pi3-pi3_transform1} using \eqref{eq:trace_formula}, we obtain
\begin{align}
&\Gamma_{\pi_{3}} (p \rightarrow 0) 
\notag \\
=& 
- \frac{2}{G_{S}} - 4 \int \frac{\dd^{4} k}{i (2 \pi)^{4}} 
\notag \\ 
& \hspace{0pt} \times 
 \frac{24 (k^{2} - 4 \sigma_{0}^{2} - 16 t_{0}^{2}) }{ (k^{2} - 4 \sigma_{0}^{2} - 16 t_{0}^{2})^{2} - 64t_{0}^{2} (k_{1}^{2} + k_{2}^{2} + 4 \sigma_{0}^{2}) }
\notag \\ 
=&
- \frac{2}{G_{S}}\biggl(
1 + 48 G_{S} 
\notag \\ 
& \hspace{0pt} \times 
\int \frac{\dd^{4} k}{i (2 \pi)^{4}} 
 \frac{(k^{2} - 4 \sigma_{0}^{2} + 16 t_{0}^{2}) - 32 t_{0}^{2} }{ (k^{2} - 4 \sigma_{0}^{2} - 16 t_{0}^{2})^{2} - 64t_{0}^{2} (k_{1}^{2} + k_{2}^{2} + 4 \sigma_{0}^{2}) }
 \biggr)
\notag \\ 
=&
96 \int \frac{\dd^{4} k}{i (2 \pi)^{4}} 
 \frac{ 32 t_{0}^{2} }{ (k^{2} - 4 \sigma_{0}^{2} - 16 t_{0}^{2})^{2} - 64t_{0}^{2} (k_{1}^{2} + k_{2}^{2} + 4 \sigma_{0}^{2}) }. 
\label{eq:tpvf_pi3-pi3_transform2}
\end{align}
This is equivalent to the last line of Eq. \eqref{eq:tpvf_p0_pi3}. 
Note that in the final equality, we used the chiral condensation gap equation \eqref{eq:gap_chiral_coexist}.

Similarly, \eqref{eq:tpvf_p0_pi3-varpi} and \eqref{eq:tpvf_p0_varpi} can be demonstrated.

%%%%%%%%%%%%%%%%%%%%%%%%%%%%%%%%%%%%%
%%%%%%%%%%%%%%%%%%%%%%%%%%%%%%%%%%%%%
\section{\label{sec:appendixD} The identity of $su_{A}(2)$ chrial rotation and translation}

From \eqref{eq:SSB_eg_2sym2cond}, we obtain 
\begin{align}
e^{i\theta Q_A^3}
\begin{pmatrix}
{\bar \psi}\psi \\ {\bar\psi}i\gamma_5\tau^3\psi
\end{pmatrix}
e^{-i\theta Q_A^3}
&=
\begin{pmatrix}
\cos \theta &  \sin \theta \\ 
-\sin \theta & \cos \theta
\end{pmatrix}
\begin{pmatrix}
{\bar \psi}\psi \\ {\bar\psi}i\gamma_5\tau^3\psi
\end{pmatrix}
\label{eq:D1}
\end{align}
By taking the vacuum expectation values, 
\begin{align}
e^{i\theta Q_A^3}
\begin{pmatrix}
\Delta\cos qz \\ \Delta\sin qz
\end{pmatrix}
e^{-i\theta Q_A^3}
&=
\begin{pmatrix}
\cos \theta &  \sin \theta \\ 
-\sin \theta & \cos \theta
\end{pmatrix}
\begin{pmatrix}
\Delta\cos qz \\ \Delta \sin qz
\end{pmatrix}\nonumber\\
&=\begin{pmatrix}
\Delta\cos (qz-\theta) \\ \Delta\sin (qz-\theta)
\end{pmatrix}
\label{eq:D2}
\end{align}
is obtained.
Similarly, from \eqref{eq:SSB_eg_2sym2cond}, we obtain 
\begin{align}
e^{iaP_3}
\begin{pmatrix}
{\bar \psi}\psi(z) \\ {\bar\psi}i\gamma_5\tau^3\psi(z)
\end{pmatrix}
e^{-iaP_3}
&=
\begin{pmatrix}
{\bar \psi}\psi(z+a) \\ {\bar\psi}i\gamma_5\tau^3\psi(z+a)
\end{pmatrix}.
\label{eq:D3}
\end{align}
By taking the vacuum expectation values, 
\begin{align}
e^{iaP_3}
\begin{pmatrix}
\Delta \cos qz \\ \Delta\sin qz
\end{pmatrix}
e^{-iaP_3}
&=
\begin{pmatrix}
\Delta\cos (q(z+a)) \\ \Delta\sin (q(z+a))
\end{pmatrix}
\label{eq:D4}
\end{align}
is obtained.
Therefore, if we adopt $a$ as $-\theta/q$, then
\begin{align}
&e^{i\left(-\theta/q\right)P_3}
\begin{pmatrix}
\Delta \cos qz \\ \Delta\sin qz
\end{pmatrix}
e^{-i\left(-\theta/q\right)P_3}
=
\begin{pmatrix}
\Delta\cos (qz-\theta) \\ \Delta\sin (qz-\theta)
\end{pmatrix}\nonumber\\
&=
e^{i\theta Q_A^3}
\begin{pmatrix}
\Delta\cos qz \\ \Delta\sin qz
\end{pmatrix}
e^{-i\theta Q_A^3}
\label{eq:D5}
\end{align}
is obtained. 
Thus, the chiral rotation due to $Q_A^3$ and the translation $-P_3/q$ cause the same transformation. 
This is the reason why the dimensional parameter $s$ should be adpted as $-1/q$ in Sec. \ref{sec:level4-3}.

%%%%%%%%%%%%%%%%%%%%%%%%%%%%%%%%%%%%%
%%%%%%%%%%%%%%%%%%%%%%%%%%%%%%%%%%%%%
\section{\label{sec:appendixC}The case of a single symmetry breaking by three or four field operators}
Now, we consider the case where a single generator $Q^{A}$ is broken by three field operators, $\Phi_{i}, \Phi_{j}$, and $\Phi_{k}$: 
\begin{align}
	\begin{split}
\langle [i Q^{A}, \Phi_{i}] \rangle &= v_{\alpha} \neq 0, 
\\ 
\langle [i Q^{A}, \Phi_{j}] \rangle &= v_{\beta} \neq 0, 
\\ 
\langle [i Q^{A}, \Phi_{k}] \rangle &= v_{\gamma} \neq 0. 
	\end{split}
	\label{eq:VEV_comm_gen_1by3}
\end{align}
Also, we assume that the fields $\Phi_{i}$, $\Phi_{j}$, and $\Phi_{k}$ are orthogonal. 
Next, using $\Phi_{i}$, $\Phi_{i}$, and $\Phi_{k}$, we redefine the fields $\Phi_{1}'$ , $\Phi_{2}'$, and $\Phi_{3}'$ as follows: 
\begin{equation}
\begin{pmatrix}
\Phi_{1}' \\ \Phi_{2}' \\ \Phi_{3}' \\ 
\end{pmatrix}
=
R_{x} R_{y} R_{z}
\begin{pmatrix}
\Phi_{i} \\ \Phi_{j} \\ \Phi_{k}
\end{pmatrix}. 
\label{eq:Field_rotation_gen_1by3}
\end{equation}
Here, 
\begin{align}
R_{x} = 
\begin{pmatrix}
1 & 0 & 0  \\ 
0 & \cos \theta_{x} & - \sin \theta_{x} \\ 
0 & \sin \theta_{x} & \cos \theta_{x}
\end{pmatrix}, 
\\ 
R_{y} = 
\begin{pmatrix}
\cos \theta_{y} & 0  & \sin \theta_{y} \\ 
0 & 1 & 0 \\ 
- \sin \theta_{y} & 0  & \cos \theta_{y}
\end{pmatrix}, 
\\ 
R_{z} = 
\begin{pmatrix}
\cos \theta_{z} & - \sin \theta_{z} & 0 \\ 
\sin \theta_{z} & \cos \theta_{z} & 0 \\ 
0 & 0 & 1
\end{pmatrix}. 
\end{align}
Since there is one broken generator $Q^{A}$, by appropriately rotating, we can obtain one NG mode and two Higgs modes. As these modes are orthogonal to each other, the Higgs modes should have an arbitrary rotation parameter within the plane defined by the NG mode. Therefore, if we consider $\theta_{x}$ as an arbitrary parameter, $\Phi_{1}'$ without $\theta_{x}$ can be considered as $\Phi_{\mathrm{NG}}$: 
\begin{align}
\Phi_{\mathrm{NG}} = \Phi_{1}' 
&= 
\cos \theta_{y} \cos \theta_{z} \Phi_{i}  - \cos \theta_{y} \sin \theta_{z} \Phi_{j} + \sin \theta_{y} \Phi_{k}, 
\\ 
\Phi_{\mathrm{H}_{1}} = \Phi_{2}' 
&=
(\sin \theta_{x} \sin \theta_{y} \cos \theta_{z} + \cos \theta_{x} \sin \theta_{z}) \Phi_{i} 
\notag \\
&+ (- \sin \theta_{x} \sin \theta_{y} \sin \theta_{z} + \cos \theta_{x} \cos \theta_{z}) \Phi_{j}
\notag \\ 
&- \sin \theta_{x} \cos \theta_{y} \Phi_{k}, 
\\ 
\Phi_{\mathrm{H}_{2}}  = \Phi_{3}' 
&=
(\cos \theta_{x} \sin \theta_{y} \cos \theta_{z} + \sin \theta_{x} \sin \theta_{z} )\Phi_{i}
\notag \\ 
&+ (- \cos \theta_{x} \sin \theta_{y} \sin \theta_{z} + \sin \theta_{x} \cos \theta_{z}) \Phi_{j}
\notag \\ 
&- \cos \theta_{x} \cos \theta_{y} \Phi_{k}. 
\end{align}
Assuming that the magnitude of the NG mode condensate is $v = \sqrt{v_{\alpha}^{2} + v_{\beta}^{2} + v_{\gamma}^{2}}$, by analogy with the case of two field operators, we have
\begin{equation}
\Phi_{\mathrm{NG}} = \frac{v_{\alpha}}{v} \Phi_{i} + \frac{v_{\beta}}{v} \Phi_{j} + \frac{v_{\gamma}}{v} \Phi_{k}, 
\end{equation} 
\begin{equation}
\langle [i Q^{A}, \Phi_{\mathrm{NG}}] \rangle = v
\end{equation} 
satisfied. In this case,
\begin{equation}
\cos \theta_{y} \cos \theta_{z} = \frac{v_{\alpha}}{v}, 
- \cos \theta_{y} \sin \theta_{z} = \frac{v_{\beta}}{v}, 
\sin \theta_{y} = \frac{v_{\gamma}}{v}
\end{equation}
hold. 
Given these conditions, $\theta_{y}$ and $\theta_{z}$ are uniquely determined. 
Using this, we obtain 
\begin{align}
&\langle [i Q^{A}, \Phi_{\mathrm{H}_{1}}] \rangle \notag \\
&=
v[ \cos \theta_{y} \cos \theta_{z} (\sin \theta_{x} \sin \theta_{y} \cos \theta_{z} + \cos \theta_{x} \sin \theta_{z})  
\notag \\
&\hspace{4pt}
- \cos \theta_{y} \sin \theta_{z} (- \sin \theta_{x} \sin \theta_{y} \sin \theta_{z} + \cos \theta_{x} \cos \theta_{z})
\notag \\ 
&\hspace{8pt}
+ \sin \theta_{y} (- \sin \theta_{x} \cos \theta_{y}) ]
\notag \\ 
&= 0, 
\end{align}
\begin{align}
&\langle [i Q^{A}, \Phi_{\mathrm{H}_{2}}] \rangle \notag \\
&=
v[ \cos \theta_{y} \cos \theta_{z} (\cos \theta_{x} \sin \theta_{y} \cos \theta_{z} + \sin \theta_{x} \sin \theta_{z} )
\notag \\ 
&\hspace{4pt}
- \cos \theta_{y} \sin \theta_{z} (- \cos \theta_{x} \sin \theta_{y} \sin \theta_{z} + \sin \theta_{x} \cos \theta_{z})
\notag \\ 
&\hspace{8pt}
- \sin \theta_{y} (\cos \theta_{x} \cos \theta_{y})] 
\notag \\ 
&= 0. 
\end{align}

Next, we consider the case where a single generator $Q^{A}$ is broken by four field operators, $\Phi_{i}, \Phi_{j}$, $\Phi_{k}$, and $\Phi_{l}$: 
\begin{align}
	\begin{split}
\langle [i Q^{A}, \Phi_{i}] \rangle &= v_{\alpha} \neq 0, 
\\ 
\langle [i Q^{A}, \Phi_{j}] \rangle &= v_{\beta} \neq 0, 
\\ 
\langle [i Q^{A}, \Phi_{k}] \rangle &= v_{\gamma} \neq 0, 
\\ 
\langle [i Q^{A}, \Phi_{l}] \rangle &= v_{\delta} \neq 0. 
	\end{split}
	\label{eq:VEV_comm_gen_1by3}
\end{align}
Also, we assume that the fields $\Phi_{i}$, $\Phi_{i}, \Phi_{k}$, and $\Phi_{l}$ are orthogonal. 
Now, using $\Phi_{i}$, $\Phi_{j}$, $\Phi_{k}$, and $\Phi_{l}$, we redefine the fields $\Phi_{1}'$, $\Phi_{2}'$, $\Phi_{3}'$, and $\Phi_{4}'$ as follows: 
\begin{equation}
\begin{pmatrix}
\Phi_{1}' \\ \Phi_{2}' \\ \Phi_{3}' \\ \Phi_{4}' \\ 
\end{pmatrix}
=
R_{12} R_{13} R_{14} R_{23} R_{24} R_{34}
\begin{pmatrix}
\Phi_{i} \\ \Phi_{j} \\ \Phi_{k} \\ \Phi_{l}
\end{pmatrix}, 
\label{eq:Field_rotation_gen_1by3}
\end{equation}
where
\begin{equation}
R_{12} = 
\begin{pmatrix}
C_{12} & - S_{12} & 0 & 0 \\ 
S_{12} & C_{12} & 0 & 0 \\ 
0 & 0 & 1 & 0 \\ 
0 & 0 & 0 & 1
\end{pmatrix}, 
R_{13} = 
\begin{pmatrix}
C_{13} & 0 & - S_{13} & 0 \\ 
0 & 1 & 0 & 0 \\ 
S_{13} & 0 &  C_{13} & 0 \\ 
0 & 0 & 0 & 1
\end{pmatrix},
\end{equation}
\begin{equation}
R_{14} = 
\begin{pmatrix}
C_{14} & 0 & 0 & - S_{14} \\ 
0 & 1 & 0 & 0 \\ 
0 & 0 & 1 & 0 \\ 
S_{14} & 0 & 0 & C_{14}
\end{pmatrix}, 
R_{23} = 
\begin{pmatrix}
1 & 0 & 0 & 0 \\ 
0 & C_{23} & - S_{23} & 0 \\ 
0 & S_{23} & C_{23} & 0 \\ 
0 & 0 & 0 & 1
\end{pmatrix}, 
\end{equation}
\begin{equation}
R_{24} = 
\begin{pmatrix}
1 & 0 & 0 & 0 \\ 
0 & C_{24} & 0 & - S_{24} \\ 
0 & 0 & 1 & 0 \\ 
0 & S_{24} & 0 & C_{24}
\end{pmatrix}, 
R_{34} = 
\begin{pmatrix}
1 & 0 & 0 & 0 \\ 
0 & 1 & 0 & 0 \\ 
0 & 0 & C_{34} & - S_{34} \\ 
0 & 0 & S_{34} & C_{34} 
\end{pmatrix}. 
\end{equation}
Here, $S_{ij} = \sin \theta_{ij}$ and $C_{ij} = \cos \theta_{ij}$, and $\theta_{ij}$ with $(i, j) = (1,2), (1, 3), (1, 4), (2, 3), (2, 4), (3, 4)$ denote the rotation parameters.

Since there is one broken generator $Q^{A}$, by appropriately rotating, we can obtain one NG mode and three Higgs modes. As these modes are orthogonal to each other, the Higgs modes should have three arbitrary rotation parameters within the three-dimentional space defined by the NG mode. 
Since
\begin{align}
\Phi_{4}' 
= 
S_{14} \Phi_{1} + C_{14} S_{24} \Phi_{2} + C_{14} C_{24} S_{34} \Phi_{3} + C_{14} C_{24} C_{34} \Phi_{4}, 
\end{align}
we can consider $\Phi_{4}'$ as the NG mode $\Phi_{\mathrm{NG}}$, and $\Phi_{1}', \Phi_{2}', \Phi_{3}'$ as the Higgs modes $\Phi_{\mathrm{H}_{1}}, \Phi_{\mathrm{H}_{2}}, \Phi_{\mathrm{H}_{3}}$. Consequently, $\theta_{12}$, $\theta_{13}$ and $\theta_{23}$ become arbitrary parameters, consistent with the above discussion. 
Assuming that the magnitude of the NG mode condensate is $v = \sqrt{v_{\alpha}^{2} + v_{\beta}^{2} + v_{\gamma}^{2} + v_{\delta}^{2} }$, by analogy with the case of two and three field operators, we have
\begin{align}
&\Phi_{\mathrm{NG}} = \frac{v_{\alpha}}{v} \Phi_{i} + \frac{v_{\beta}}{v} \Phi_{j} + \frac{v_{\gamma}}{v} \Phi_{k} + \frac{v_{\delta}}{v} \Phi_{l}, 
\\
&\langle [i Q^{A}, \Phi_{\mathrm{NG}}] \rangle = v
\end{align} 
satisfied. In this case,
\begin{align}
&\sin \theta_{14} = \frac{v_{\alpha}}{v}, 
\qquad
\cos \theta_{14} \sin \theta_{24} = \frac{v_{\beta}}{v}, 
\nonumber\\ 
&\cos \theta_{14} \cos \theta_{24} \sin \theta_{34} = \frac{v_{\gamma}}{v}, 
\qquad 
\cos \theta_{14} \cos \theta_{24} \cos \theta_{34} = \frac{v_{\delta}}{v}
\end{align}
hold. 
Given these conditions, $\theta_{14}$, $\theta_{24}$, and $\theta_{34}$ are uniquely determined. 
Using this, we obtain 
\begin{align}
&\langle [i Q^{A}, \Phi_{\mathrm{H}_{1}}] \rangle
= 0, 
\\
&\langle [i Q^{A}, \Phi_{\mathrm{H}_{2}}] \rangle
= 0, 
\\
&\langle [i Q^{A}, \Phi_{\mathrm{H}_{3}}] \rangle
= 0. 
\end{align}

%%%%%%%%%%%%%%%%%%%%%%%%%%%%%%%%%%%%%
%%%%%%%%%%%%%%%%%%%%%%%%%%%%%%%%%%%%%
%-----reference-----
 
\end{document}